\documentclass[11pt]{article}
\usepackage{geometry}                % See geometry.pdf to learn the layout
\usepackage{setspace}
\usepackage{graphicx}
\usepackage{amssymb}
\usepackage{epstopdf}
\usepackage{multirow}
\usepackage{extarrows}
\usepackage{endnotes}
\title{Symmetries and Paraparticles as a Motivation for Structuralism}
\author{Adam Caulton and Jeremy Butterfield}
\date{}                                           % Activate to display a
\begin{document}
\maketitle

\begin{abstract}
This paper develops an analogy proposed by Stachel between general
relativity (GR) and quantum mechanics (QM) as regards permutation
invariance. Our main idea is to overcome Pooley's criticism of the analogy
by appeal to paraparticles.

In GR the equations are (the solution space is) invariant under
diffeomorphisms permuting spacetime points. Similarly, in QM the equations
are invariant under particle permutations. Stachel argued that this
feature---a theory's ``not caring which point, or particle, is which"---supported
a structuralist ontology.

Pooley criticizes this analogy: in QM  the (anti-)symmetrization of fermions
and bosons implies that each individual state (solution) is fixed by each
permutation, while in GR a diffeomorphism yields in general a distinct,
albeit isomorphic, solution.

We define various versions of structuralism, and go on to formulate Stachel's and Pooley's positions, admittedly in our own terms.  We then reply to Pooley. Though he is right about fermions and bosons, QM
equally allows more general types of symmetry, in which states (vectors,
rays or density operators) are {\em not} fixed by all permutations (called
`paraparticle states'). Thus Stachel's analogy is revived.
\end{abstract}

\tableofcontents

\section{Introduction}

In this paper we use the idea of paraparticles\footnote{Following Messiah and Greenberg [1964], we use the term `paraparticle' to denote particles of \emph{any} symmetry type other than boson and fermion.  This terminology is not universal: Huggett [2003], inspired by Hartle, Stolt \& Taylor [1970], associates the term `paraparticle' with a restricted class of symmetry types; his term `quarticle' is closer to our `paraparticle'.  (A term yet more inclusive than our `paraparticle' is Greenberg's (e.g.~[2009]) `quon', which covers states which do not even satisfy the indistinguishability postulate, but which transform as representations of the symmetric group.)}---a mathematically natural form of particle symmetry in quantum theory (albeit less familiar than Bose-Einstein and Fermi-Dirac)---to add to a recent debate between John Stachel and Oliver Pooley about whether general relativity and non-relativistic quantum mechanics provide analogous motivations for a structuralistic ontology.

In his [2002], Stachel attempted to unify Leibniz equivalence and anti-haecceitism, two interpretative positions in general relativity and quantum mechanics respectively, by arguing for a generalisable, over-arching motivation for them both.  This was his `hole argument for sets'.\footnote{We should make clear that, although Stachel's main effort is to develop this argument, his favoured reason for quantum anti-haecceitism is different: it appeals to a version of Ockham's razor due to Hertz ([2002], p.~243).}  The argument is abductive: structuralism best explains, or makes palatable, a feature of certain theories.  This feature is a special form of indeterminism: theories which suffer from it fail to pick out unique predictions amongst a certain set of isomorphic alternatives.  Structuralism dissolves this indeterminism, by saying that the difference between these alternatives is to be interpreted as an artefact of using a system of description with redundant elements.  That is: two such isomorphic alternatives represent the same physical state of affairs; so that, both physically and philosophically speaking, there is after all no indeterminism.

We concur with the consensus, including Stachel, that this is the best way to view Leibniz equivalence in general relativity.  But, as Pooley pointed out ([2006]), quantum mechanics, in the form Stachel considers it, does not have the requisite formal structure to motivate structuralism in an analogous way.  Pooley's objections echo claims that the metaphysics of identity and individuality for quantum particles is under-determined by the quantum formalism, levelled by French (e.g.~[1998]) and others.

Our main claim will be that, when we consider quantum mechanics in a more general form---specifically, allowing paraparticle states---an analogy with general relativity is rehabilitated; and so Stachel's argument for anti-haecceitism can be revived.  So while Pooley and French are right that quantum mechanics, in its usual form, provides no motivation for a structuralist ontology, we recommend, following Messiah \& Greenberg's ([1964]) advocacy of paraparticles, a more general formulation.  The alternative formulation finds a home for Stachel's claimed analogy, and his associated abductive argument for structuralism.

The plan is as follows. Section \ref{s:structure} contains a general discussion of what we are calling `structuralism', and a first look at its application to general relativity and quantum mechanics.  Throughout, our discussion of quantum theory concentrates, as do Stachel and Pooley, on quantum mechanics for fixed particle number, i.e.~we do not consider quantum field theory.  Section \ref{s:hasets} presents Stachel's generalised hole argument for sets.  Section \ref{s:fixed} discusses Pooley's reply, and the related claim that quantum physics under-determines the metaphysics of identity and individuality for quantum particles.  Section \ref{s:para} introduces paraparticles and shows how a quantum mechanics which allows them resuscitates Stachel's argument.

We admit at the outset that, according to the empirical evidence so far, no fundamental particles are paraparticles; (this is also true in quantum field theory).  But we stress that our revival of Stachel's argument will not depend on their being fundamental paraparticles; or even on the (weaker) claim that paraparticle \emph{states} occur in nature.  The revival will only need the possibility of paraparticles.

\section{Structuralism applied to general relativity and quantum mechanics}\label{s:structure}
\subsection{Structuralism and individuality}\label{subs:structure}

Structuralism, as we will understand it, comes in a variety of strengths, but the common element is a claim about the individuality of the objects we are interested in. `Individuality' is something of a term of art. We propose the following definition: an object is an individual, or has individuality, if and only if it has some property that all other objects lack (which we will sometimes call its `individuating property').  Here we intend `property' widely: familiar concerns about the expressiveness of first-order languages aside, we intend it to capture anything that might be represented in a formal language by an open formula with only one free variable.  As such, the individuating property---i.e.~the property responsible for an object's individuality---may be so convoluted and heterogeneous that no metaphysician would consider it ÔnaturalÕ.  It may be `extrinsic', in the sense that it makes ineliminable reference to \emph{other} objects, perhaps all of them.  The property may even fail to be finitely expressible (though here we run up against the standard finitary conception of a formal language, some one-place open formula of which was supposed to express the property).

According to our definition of individuality, it is orthodoxy in formal logic that every object is an individual, for $a$ and $a$ alone has the unique property of \emph{being $a$}, represented by the formula `$x=a$'.  But structuralism denies this claim.  That is, structuralism denies that objecthood is sufficient for individuality.  Its central claim is that individuality is grounded, if at all, only on \emph{qualitative} properties and relations.  The terms `qualitative' and `grounded' are slippery and obscure, but we can make `grounded' more precise: structuralism holds that any object's individuating property is a (logical) construction only of its qualitative properties, and/or of the qualitative relations it has to other objects.  As for `qualitative', we will not attempt a definition, but just refer the reader to Adams' ([1979], pp.~6-9) discussion, and trust that the notion will be sufficiently clear in our ensuing discussion.

Again, the properties and relations, which, according to a structuralist, individuate objects, may be as `unnatural' as you like---they may even refer to other objects, though only by description and not by name.  Being the tallest person in the room, for example, is a property which ineliminably refers to everyone in the room.  Thus, qualitative properties and relations, by their nature, do not identify the objects, if any, that they refer to, \emph{except} by means of the qualitative properties and relations that \emph{they} enjoy (their being shorter than someone in the room).  The property of being $a$ is therefore not qualitative: it is a `haecceitistic property'.  So according to structuralism, it cannot ground $a$'s individuality.  For the structuralist, every object's individuality, if it has it, is grounded in the pattern of instantiation---the structure---of the qualitative properties and relations in which it is embedded.\footnote{So far, our definition of structuralism has not ruled out the position that every object actually has an individuating essence, where these essences are entirely `intrinsic', qualitative properties.  Some will feel uncomfortable associating the term `structuralism' with this position, but we allow it as a limiting case.  Other structuralists might wish to add a positive claim that there \emph{are} some non-individuals, or that \emph{some} object's individuating property is extrinsic.   But for us it is sufficient that there \emph{could have} been non-individuals, or extrinsic individuating properties.  We will \emph{not} allow as structuralist, however, the claim that every object \emph{must} have an intrinsic individuating essence.  Suffice it to say, these issues do not arise for the theories of interest here, viz.~general relativity and quantum mechanics, since in neither theory are there any plausible candidates for qualitative individuating essences for the objects concerned.  For more discussion of structuralism and its rivals, cf.~Caulton \& Butterfield [2010a].\label{fn:indessences}}

In terms of a formal language, structuralism can thus be characterised as the view that no individual's uniquely applicable formula contains any names---except, perhaps, names for other individuals.  Indeed, according to structuralism, if an object can be picked out at all, then its name-free, uniquely applicable formula will do.

As we said, structuralism comes in various strengths, and we can place them on a spectrum from weak to strong.  At the weak end, the numerical diversity of objects is accepted as primitive, just as it is by the haecceitist; even though, \emph{contra} haecceitism, individuality is qualitatively grounded.  Accordingly, this weak structuralism allows objects which are `non-individuals', and even allows objects that are indiscernible from one another.  This form of structuralism is new;\footnote{The position has, or at least positions very similar to it have, been discovered and defended, apparently independently, by Ladyman ([2007b], pp.~33-37), Leitgeb \& Ladyman ([2008]) and Esfeld \& Lam ([2006]); although the core ideas are also present in Pooley ([2006], pp.~101-103), Belot \& Earman ([2001], pp.~228-229, 248) and Hoefer ([1996]), and may even be traced back to Carnap ([1928], \S\S10-16; [1966], Ch.~26) and the idea of a Ramsey sentence.  For a review of the metaphysical options, see Caulton \& Butterfield [2010a].\label{fn:sr}} and at first, perhaps, even puzzling.

Along the way to the strong end of the spectrum, we find versions of structuralism that ground diversity, as well as individuality, in the differential instantiation of qualitative properties and relations.  In doing so, these versions are committed to some version of the identity of indiscernibles.  But this may still allow for non-individuals: for two objects may be discernible by some relation though each plays the same role in the mosaic of qualitative properties and relations as the other, so that no \emph{one-place} formula distinguishes either one from the other.\footnote{This gap between qualitatively grounded diversity and qualitatively grounded individuality can be seen first in Hilbert \& Bernays' ([1934]) proposed account of identity. It was noticed there by Quine ([1960], [1970], [1976]) and has been put to use in the philosophy of physics by Saunders ([2003a], [2003b], [2006a]), Muller \& Saunders ([2008]) and Muller \& Seevinck ([2009]).  This gap between individuality and diversity is that between what Quine called \emph{absolute} and non-absolute (comprising \emph{relative} and \emph{weak}) discernment: which, roughly speaking, is discernment by the differential instantiation of, respectively, one-place and two-place open formulas.  The relations between the various kinds of discernment are discussed in Caulton \& Butterfield [2010a].\label{fn:HB}}

At the strong end of the spectrum, non-individuals are ruled out altogether, along with indiscernibles.  But, to deserve the epithet `structuralism' (cf.~footnote \ref{fn:indessences}), the diversity and/or individuality of the objects must still be grounded, or potentially grounded, in qualitative relations as opposed to intrinsic, qualitative essences.

Now that we have drawn attention to the variety of structuralist positions, we emphasise that, for the rest of this paper, our discussion of structuralism applies to the \emph{disjunction} of all the positions on this spectrum.  That is: we do not commit ourselves to \emph{which} specific strength of structuralism best suits general relativity or quantum mechanics. Nor do we commit ourselves to the view that structuralisms of all strengths suit these theories equally well.  In fact, on the contrary: we agree with the literature (e.g.~French \& Redhead [1988], Butterfield [1993]) that the strongest version of structuralism, with its strong version of the identity of indiscernibles, is incompatible with quantum mechanics---and we would say the same about general relativity.  However, this still leaves undecided whether spacetime points and quantum particles should be treated as \emph{primitively} distinct or else distinct according to some (relational) qualitative difference. (Cf.~footnotes \ref{fn:sr} and \ref{fn:HB} respectively).

\noindent\emph{Digression: the name `structuralism'}\quad
 We are using the name `structuralism' for a very specific thesis concerning the individuality, and perhaps also the numerical diversity, of objects, where `objects' are also to be understood in a precise and narrow sense, namely as in the tradition of Frege and Quine.  But in the philosophy of science \emph{alone} (not to mention Saussure \emph{et al}!)~many different views go by the name `structuralism'.  So we should be explicit about what affinities, if any, our structuralism has with those other positions or standpoints.

First of all, there are heuristic links with structural realism, taken as the claim that the objects of study in the physical sciences (`objects' now in the loose sense) can be known only---due either to an insuperable epistemic gulf, or the sheer absence of anything else to know---in terms of their structure---meaning, roughly, the mathematics that governs them.  We have in mind here the structural realism of Worrall [1989] and Ladyman [1998].  Particularly worthy of note is what Ladyman calls `epistemic structural realism', and kindred positions such as `Ramsey-sentence realism' (cf.~Ketland [2004], Melia \& Saatsi [2006], Ainsworth [2009]).  The idea is to conjoin a theory's sentences, then replace its names (and perhaps also its predicate symbols) with free variables, and then existentially quantify over the variables (perhaps also adding a statement of uniqueness for each).  The resulting sentence is then considered to be an exhaustive description of the theory's subject matter.  This is certainly in the spirit of structuralism: the entities associated with the existentially bound variables are conceived of functionally, i.e.~in terms of the role they play in a structure.  Our structuralism follows this in spirit.\footnote{Our structuralism goes some way towards Carnap's ([1928]) vision of abstracting a theory's content until nothing is left but bare nodes and arrows; for, according to structuralism, an object is nothing but a node in the web of properties and relations.  But from Carnap's point of view, our structuralism is only half-hearted, since it makes no analogous commitment to ``abstracting from" the properties and relations, i.e.~defining them in terms of which objects enjoy them.  In that sense our structuralism is quidditist, though anti-haecceitist.  (The term `quidditism' seems to be due to Armstrong ([1989], p.~59), in analogy with haecceitism.)}

But we would like to emphasise a distinction between inter- and intra-theoretic structuralism.  Structuralism, as we are using it, may be the best interpretative position regarding the objects of a single \emph{given} theory---indeed, we will argue that this is so for both general relativity and quantum mechanics. But this does not in the slightest imply structural \emph{realism}: the view that, by eschewing all but the mathematics in our understanding of theoretical entities, we may---according to Worrall---secure the survival of successfully referring theoretical terms through (perhaps violent) theory-change, or---according to Ladyman---overcome the under-determination by physical formalism of interpretative metaphysics.  On the contrary, our structuralism is best seen as one interpretative strategy amongst many, for a given theory.  It is a strategy with no guarantee of surviving theory-change; and, as regards putative under-determination, since our structuralism is one interpretative strategy among others, it is---we admit!---one of the rival views in contention, not a solution that somehow dissolves the contest between them.\footnote{Pooley ([2006], p.~91, parag.~3) makes this last point, in the course of a fine discussion of the contrast between the inter-theoretic and intra-theoretic meanings of structuralism (p.~86-99). Also, beware: Esfeld \& Lam ([2006]) call their position `moderate structural realism', suggesting the same inter-theoretic ambitions as Worrall and Ladyman; but their position is in fact a structuralism in our, \emph{intra}-theoretic, sense, viz. the weakest kind, since they take objects' diversity as primitive; cf.~footnote \ref{fn:sr}.}
\emph{End of digression.}

\subsection{The semantics of the structuralist}\label{subs:semstr}
Grounding individuality qualitatively has consequences for the specification of possible states of affairs.  In particular, we cannot make sense of what Lewis ([1986], p.~221) called `haecceitistic differences': differences to do with which object occupies which role in the mosaic of qualitative matters of fact.  Since an object's individuality is provided, if it all, by the role that it plays in the structure of which it is a part, there simply is nothing else to grasp it with, that could be used to give sense to, for example, the object's swapping roles with another.  A structuralist simply does not agree with the haecceitist that a permutation of objects ``underneath" the mosaic makes any sense.

But here the structuralist runs up against an aspect of model theory (formal semantics) which threatens to make her position hard to state.  Model theory allows without demur that an object may be individuated independently of its qualitative role in a structure.\footnote{This point is often emphasised in the philosophy of modality; a good early example is Kaplan [1966].} Or think of Lewis's ([1986], p.~145) `Lagadonian' languages, in which objects are their own names.  In philosophical terms, this trick ensures that haecceities are always at hand to give sense to a permutation of ``bare" objects underneath the qualitative properties and relations.

So, not wishing to cause too much of a stir, our structuralist might express her position in a way more generous to the orthodox practice of model theory, with its allowance for haecceities.  We endorse such indulgence. So from now on we shall imagine that even our structuralist accepts, for the sake of retaining standard model theory, that permutations of `bare' objects do make sense.  More precisely, the structuralist will admit that such permutations typically induce a different formal representation of a state of affairs, for example, a different model in the sense of model theory. But she then expresses her position by saying that such a permutation does not produce a representation of a \emph{distinct} state of affairs.

On such a view, it is misleading to say that \emph{models}\footnote{We intend `model' to carry no connotation of satisfying a set of sentences; we avoid the alternative, more common synonyms, `structure' and `interpretation', for obvious reasons!\label{fn:models}} represent possible states of affairs, since the representation relation is not one-to-one but many-to-one.  Rather, a single state of affairs is represented by the \emph{equivalence class} of mutually isomorphic models, the isomorphism being given by a permutation of the `bare' objects in the model's domain.  This ascent from models to equivalence classes will be a major theme in the sequel.  For it suits a theory which seems ``not to care" how its objects are arranged under the mosaic of properties and relations, so long as the same qualitative pattern is instantiated, so that such theories support a structuralist intepretation.\footnote{One might object, to representing a state of affairs with an equivalence class, that it does not have the right kind of structure, whereas its elements---the models---do, albeit in virtue of what is common to all of them.  (Many thanks to Oliver Pooley for this point.)  But we will continue to advocate the equivalence class.  In this we follow the a long tradition; for example, Frege defined the extension of the concept of the number 4 as the (infinite!) set of all quartets, and did not say that each quartet equally represents the number 4.}

Stachel's contention is that general relativity and quantum mechanics are both examples of these structuralism-friendly theories; and he urges a structuralistic ontology of, respectively, spacetime points and elementary particles.  Pooley disagrees: he points out that quantum mechanics, in its usual form, is crucially disanalogous to general relativity. It is only with general relativity, he argues, that structuralism can boast any interpretative benefit.  We will see that Pooley is right in this---for the formulation of quantum mechanics that both he and Stachel consider.  But we will argue that matters change---that is, Stachel's analogy is rehabilitated---when we consider quantum mechanics in a more general setting; specifically, when we drop a usual but unobligatory restriction on states---viz.~the symmetrisation postulate---and thereby allow paraparticles.  Before that, however, we will first consider in Sections \ref{subs:GR}-\ref{subs:QM} the two theories as they appear in both Pooley and Stachel's arguments.

A final incidental remark:
we agree that the structuralist may try to be true to her original prohibition against permutations making sense.  That is: she may try to move to a different formalism in which an equivalence class of isomorphic representations is replaced by a single item, so that indeed permutations do not even make a \emph{formal} difference, or else cannot be made sense of.  In classical mechanics, this formalism goes by the name \emph{reduced state-space}, and the process of moving to it involves \emph{quotienting} the original theory's state-space (cf.~e.g.~Belot ([2003], \S 5) and Butterfield ([2006], \S\S~2.3, 7)).  In quantum mechanics too, after the structuralist has embraced paraparticles and given up the idea that each ray represents a distinct quantum state (see later), she can, if she wishes to, move to a `reduced' Hilbert space of fewer dimensions, in which a one-to-one correspondence between rays and states is regained, but permutations can no longer even be represented (cf.~Hartle \& Taylor ([1969], p.~2047)).  We emphasise that the new structuralist formalism retains formal semantics, and does not require a revised logic.\footnote{An example of such a logic is `quasi-set theory' (cf.~Krause [1992], Dalla Chiara, Giuntini \& Krause [1998], French \& Krause ([2006], ch.~7) and da Costa \& Krause [2007]), in which sentences stating identity or non-identity of objects which are (in our sense) non-individuals are not well-formed.}

\subsection{General relativity and Leibniz equivalence}\label{subs:GR}

In the philosophy of spacetime physics, the question whether a permutation of spacetime points makes a physical difference, is well established, and goes under the buzz-words `diffeomorphism covariance', and `hole argument'.\footnote{A selection of the early literature: Earman \& Norton [1987], Butterfield [1987], Rynasiewicz [1994].}   The correct answer to this question (at least, by our lights!)~is almost as well established---in favour of structuralism, i.e.~that a permutation of spacetime points does \emph{not} make a physical difference (buzz-word: `Leibniz equivalence').  But we will here introduce the issue at a leisurely pace, because it is essential to our argument, and there is plenty of opportunity for confusion.  Indeed, there is more opportunity for confusion here than in most re-tellings of the story, which normally only distinguish between the ideas of passive and active diffeomorphisms.  But here we will also need to distinguish between \emph{two} different descriptions of an \emph{active} diffeomorphism, in order to stick as closely as possible to the formal apparatus in Stachel's and Pooley's discussions.

In general relativity (GR), complete physical states are represented by what we will call \emph{histories}.
We will take a history as an ordered triple $\langle M, g, T\rangle$.\footnote{We thank Erik Curiel for the observation that the ordered \emph{pair} $\langle M,g \rangle$ suffices to pick out a unique history, since the Einstein field equations fix a unique $T$-field  given a specification for the $g$-field.  We nevertheless retain both, to reflect our view that the two fields are separate objects (for a sympathetic voice cf.~Lehmkuhl [2008]); it may also be taken to insinuate a commitment to some distinction between the kinematically and dynamically possible worlds.} $M$ is a differentiable manifold made up of points $p \in M$, equipped with four coordinate functions $x^\mu: U \subset M \to \mathbb{R},\ \mu \in \{0,1,2,3\}$, which give each point $p$ in a local `patch' $U$ its unique four-dimensional coordinates $x^\mu(p)$.\footnote{Of course, the manifold $M$ in general needs many charts $x^\mu: U \subset M \to \mathbb{R}^4$.  Care is required about the overlapping of patches; for details see e.g.~Bishop \& Goldberg ([1968], pp.~19-36).}  The metric $g$ and matter field(s) $T$ each assign to every point $p$ of the manifold a rank-2 symmetric tensor acting on the tangent space $T_pM$ at $p$.

\begin{figure}[h] 
\setlength{\unitlength}{1mm}  % selecting unit length 
\centering      % used for centering Figure 
\begin{picture}(60,50)   % picture environment with the size (dimensions) 
      % 32 length units wide, and 15 units high. 
\put(0,0){\line(1,4){5}} \put(0,0){\line(1,0){50}} \put(50,0){\line(-1,4){5}} \put(5,20){\line(1,0){40}}
\put(0,32){\framebox(20,16){}}, \put(27,32){\framebox(20,16){}} 
\put(10,13){\vector(0,1){25}} \put(9,13){\line(1,0){2}} 
\put(40,14){\vector(0,1){21}} \put(39,14){\line(1,0){2}} 
\put(12,13){\vector(1,1){22}} \put(11,14){\line(1,-1){2}} 
\thicklines
\put(12,11){\vector(1,0){24}} \put(12,10){\line(0,1){2}} 
%\put(11,26){\vector(1,0){13}}  \put(11,25){\line(0,1){2}} 
\put(50,6) {$M$} \put(50,31) {$\mathcal{T}^0_2(T_{d(p)}M)$} \put(-19,31) {${\mathcal{T}^0_2(T_pM)}$}
\put(9,10) {$p$} \put(37,10) {$d(p)$} \put(23,7){$d$} %\put(16,27.5){$d^*$}
\put(7,25){$g$} \put(41,25){$g$} \put(27,25){$d^*g$}
\put(5,40) {$g(p)$}\put(30,42) {$d^*g(p) =$} \put(30,37){$(g\circ d)(p)$} 
\end{picture} 
\caption{A diffeomorphism $d$ and its pullback $d^*$.} % title of the Figure 
\label{fig:diff}    % label to refer figure in text 
\end{figure}
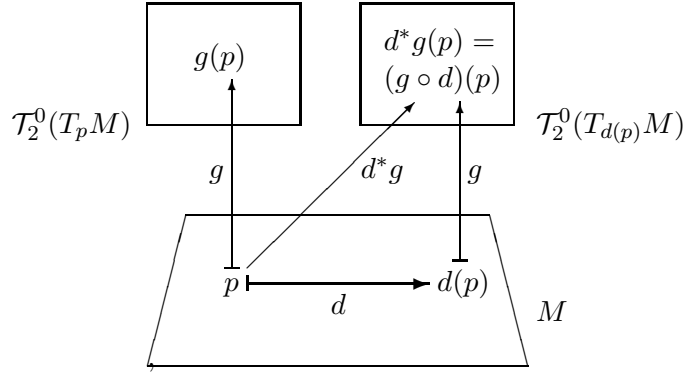

Now let us consider a \emph{diffeomorphism} $d: M \to M$; cf.~Figure \ref{fig:diff}.  To side-step potential confusion (in particular between `active' and `passive' transformations), we will not employ coordinates for the manifold points. A diffeomorphism is a ``nice" kind of reshuffling of the points of $M$. To be precise: it is a smooth bijection with a smooth inverse, so that it preserves $M$'s differential and topological structure.

The point $p$ gets mapped to the point $d(p)$. Thus $p$ takes on the role previously played by the point $d(p)$, in the sense that one can think of, for example, the field $g$ at point $p$ altering its value from $g(p)$ to $g\left(d(p)\right) \equiv \left(g \circ d\right)(p)$.  Therefore the diffeomorphism $d$ induces a transformation $d^*$ on the space of physical fields, known as the \emph{pullback by} $d$, where $d^*: g \mapsto d^*g := g \circ d$ (and similarly for $T$); cf.~Figure \ref{fig:diff}.  This (typically) generates a new history  $\langle M, g, T\rangle \mapsto \langle M, d^*g, d^*T\rangle$.  We emphasise that this history is new in the sense of being \emph{mathematically} distinct from the original; no claim is being made about any \emph{physical} difference between the histories.  We now define the pullback in a form that is more appropriate for our later discussion (Section \ref{subs:perms}):
\begin{center}
(\emph{Def $d^*g$}) \qquad
For all $p \in M$:  $p$ has the $g$-field value $d^*g(p)$ iff
the point $d^{-1}(p)$ has the $g$-field value $g(p)$ .
\end{center}
(We could similarly define (Def $d^*T$).)

Now, the dynamical equations of GR, the Einstein field equations, which relate (the values and various derivatives of) the function $g$ to the function $T$, have a feature known as \emph{diffeomorphism covariance.}  This means that, if some history $\langle M, g, T \rangle$ is a \emph{solution} to the equations, then so is every history diffeomorphically related to it, i.e.~every history of the form $\langle M, d^*g, d^*T \rangle$, where $d$ is a diffeomorphism.

This situation can be framed particularly vividly if we suppose that $d$ acts as the identity map on the spacetime in the past of a given time-slice, but smoothly diverges from it---as wildly as we like---later on.  The equations of GR then seem to be radically indeterministic: for the field values prior to any given time-slice fail to determine the values of the fields after that time slice.  For any time, the later evolution of the fields may lie anywhere within a vertiginously large space of possible trajectories, a space delimited only by the weak constraint that $d$ be smooth with smooth inverse.

This apparent indeterminism was a great worry to Einstein in 1913, when he was still struggling to formulate GR.\footnote{See e.g.~Norton [1984] for a history.}  Nowadays there is almost universal consensus\footnote{The consensus is stronger in the physics literature than in the philosophy literature.  For the physics consensus, see e.g.~Wald ([1984], pp.~253-5, 260, 266), and Hawking \& Ellis ([1973], pp.~227-8, 248).  The classic first statement of the philosophers' consensus is Earman \& Norton [1987].  For early dissenting responses, see Butterfield [1987] and Maudlin [1990].} that the correct response is to consider diffeomorphically related histories as representatives of the \emph{same} physical state of affairs.  In the philosophical literature, this position is usually known as \emph{Leibniz equivalence},\footnote{The term originates in Earman \& Norton [1987].} in tribute to the great relationist.

Diffeomorphisms are, of course, a special kind of permutation.  So if (as is customary) we consider GR as a theory whose objects are manifold points, then GR is a theory that does not care which is which, in exactly the sense of Section \ref{subs:structure} and Section \ref{subs:semstr}.  Structuralism, in the form of Leibniz equivalence, prevents this indifference from resulting in under-determination of \emph{physical} predictions; in particular it avoids a radical physical indeterminism.

\subsection{Quantum mechanics and anti-haecceitism}\label{subs:QM}
In the quantum mechanics of many-particle systems (QM),\footnote{As already mentioned, we discuss only quantum mechanics for fixed particle number.} the topic of object permutations is less settled than in GR.  Anti-haecceitism---the term we will use to label structuralism about quantum particles, and which, following Stachel, we intend to make analogous with Leibniz equivalence---is for some authors the most natural intepretative option.\footnote{E.g.~Redhead \& Teller [1992] and Saunders [2003a].}  Haecceitism appears to have little or no support.  On the other hand, many are the voices who urge that all the scientific and philosophical evidence leaves one with under-determination between these two views.\footnote{E.g.~French \& Redhead ([1988], pp.~237-238), French [1998], French \& Krause ([2006], pp.~144-150), French \& Ladyman [2003]. A more radical view denies that particles are even the objects of quantum mechanics in the first place, but we will not discuss this view here.\label{manyvoices}}  

Here we will present anti-haecceitism on analogy with the Leibniz equivalence of the previous Section.   For the two metaphysical views to be analogous in the way Stachel wanted, there should be some quandary issuing from the formalism, which, on analogy with diffeomorphism covariance, gives an interpretative advantage to anti-haecceitism.  This quandary should relate to the under-determination of mutually isomorphic models by the theory's equations, or in some other way suggest that the theory ``does not care" which of the theory's objects---which particles---play which role in the qualitative structure of a possible world.

In fact there are two salient ways in which QM might be said not to care ``which is which".  One of these ways considers quantum statistics, in particular how they diverge from the statistics which are classically (and, it is mistakenly presumed, haecceitistically) expected.  We will argue that this way offers a false path to anti-haecceitism.  We will present it in this Section because it is familiar in the literature, and because---within the formulations considered by Stachel---it is the \emph{only} plausible quantum counterpart of diffeomorphism covariance.  (But we emphasise that he does not explicitly commit himself to this argument being a motivation for anti-haecceitism.\footnote{Stachel does not present the argument explicitly, but cites it as a standard argument for anti-haecceitism ([2002], p.~245, parag.~3), and seems sympathetic to it (p.~236, parag.~6; p.~245, parag.~4--p.~246).})  However, Stachel's discussion is given under too narrow a formulation of quantum mechanics: he assumes ([2002], fn.~76) that a certain assumption, \emph{the symmetrisation postulate}, is imposed.  But this assumption is not compulsory.  And, as we shall see (in Section \ref{s:para}), it is only when this requirement is relaxed that we obtain a genuine analogy between general relativity and quantum mechanics---between Leibniz equivalence and anti-haecceitism.

This leads to the second way in which QM might be said not to care ``which is which", which will be endorsed later (in Section \ref{s:para}).  But to anticipate: when the class of physical quantities is restricted in accordance with the \emph{indistinguishability postulate}, a specification of the expectation values for all quantities typically \emph{fails} to pick out a unique ray in Hilbert space as representing the corresponding (pure) physical state. Rather it typically picks out only a multi-dimensional subspace of the Hilbert space. We describe this as QM ``not caring which is which", because what characterises the subspace is precisely that it is spanned by rays differing only as to which single-particle state is associated with which particle label.  So anti-haecceitism is indeed the view that the \emph{entire subspace} represents a single pure physical state---in exact analogy with Leibniz equivalence.

To explain all this, we will first (Section \ref{subsubs:perms}) present the candidate for what in QM are the ``models", i.e.~the mathematical representatives of physically possible states of affairs, which are to be understood on analogy with the histories of GR.  Then (Section \ref{subsubs:SPIP}) we will discuss the idea of a particle permutation, on analogy with a diffeomorphism in GR, and spell out the two postulates of QM, the symmetrisation and indistinguishability postulates, on which our ensuing discussion will turn.  We will then be in a position to present (Section \ref{subsubs:badarg}) the fallacious argument for anti-haecceitism from the quantum statistics.

\subsubsection{Pure and mixed states; permutations}\label{subsubs:perms}

We consider an assembly of $n$ quantum particles, $p_1, p_2, \ldots p_n$, which are `indistinguishable'.  This term (as well as the unfortunate `identical') means that the particles belong to the same \emph{species}: each possesses the same mass, charge and intrinsic spin.\footnote{In the jargon of Quine ([1960], p.~230), the particles are all \emph{absolutely indiscernible} one from another; cf.~footnote \ref{fn:HB}, and its references' discussion of QM's violation of the strong version of the identity of indiscernibles.}  For simplicity we will assume each particle has a finite-dimensional Hilbert space $\mathfrak{h}_l, l \in \{1,2,\ldots n\}$ of dimension $d$.\footnote{But our entire discussion is generalisable to infinite-dimensional Hilbert spaces.}

In QM, \emph{pure} physical states are usually represented by \emph{rays}.  (The `usually' will be in force until we consider paraparticles in Section \ref{s:para}.)  We may take a ray to be defined by a projection operator $W$, onto a one-dimensional subspace of the assembly's Hilbert space $\mathcal{H} :=  \bigotimes_l^n \mathfrak{h}_l$ of $d^n$ dimensions.  So $W$ may be represented by a $d^n \times d^n$ self-adjoint, idempotent matrix with unit trace (i.e.~$W^\dag = W$, $W^2 = W$ and $Tr(W)=1$).\footnote{For the sake of a more exact later comparison with GR and Stachel's notation, we could instead take a ray as an ordered pair $\langle\{p_1, p_2, \ldots p_n\}, W \rangle$, where $p_1, p_2, \ldots p_n$ are our $n$ indistinguishable particles and $W$ is a projection operator.  We will not do this; but the reader may take the specification of the particles as implicit in $W$.}  Each $W$ projects all vectors in $\mathcal{H}$ onto a single one-dimensional subspace, defined by some unit vector $|\psi\rangle$.  Therefore $W$ may be written $W = |\psi\rangle\langle\psi|$.

We may expand the space of represented physical possibilities to include \emph{mixed states}, though we shall continue to reserve the name `ray' for pure states.  For mixed states, the operator $W$ is a more general mathematical object, a \emph{density operator}: a self-adjoint, non-negative operator with unit trace, which may be represented by a corresponding $d^n \times d^n$ matrix.  A density operator is therefore a convex combination of projection operators: $W = \sum_i w_i|\psi_i\rangle\langle\psi_i|$, where each $w_i \in \mathbb{R}$ and $\sum_i w_i = 1$. The special case of a pure state, or one-dimensional projector, corresponds to all but one of the $w_i$ being zero.  A density operator specifies expectation values for every quantity $Q$, each represented by a self-adjoint operator, by the Born rule: $\langle Q \rangle = Tr(W Q)$.\footnote{Incidentally: though time will not be much of an issue in our discussion of quantum theory, we adopt the Schr\"odinger picture, in which rays evolve over time and quantities do not. (Cf.~the digression at the end of Section \ref{subsubs:badarg}).  But we could equally well adopt the Heisenberg picture, in which it is the quantities, and not the rays, that evolve over time.\label{fn:SvsH}}

We now consider the effect of a permutation $\pi$ of the particles $p_1, p_2, \ldots p_n$, on our mathematical representatives for states.  It will suffice to consider their effect on the rays---the pure states---since any effect on them induces an effect on their various convex combinations.  Actually, it will be clearest to consider the behaviour under a permutation of the unit vector $|\psi\rangle$ in the ray for which $W = |\psi\rangle\langle\psi|$. 

For definiteness, let us take the case of three particles, $n=3$. Suppose that each particle's single-particle Hilbert space $\mathfrak{h}_l,\ l\in\{1,2,3\}$, is spanned by the unit vectors $|\xi_i\rangle$, or by the unit vectors $|\eta_j\rangle$, or by $|\zeta_k\rangle$, where $i,j,k \in \{1, 2, \ldots d\}$. Then we may write the unit vector $|\psi\rangle$, which lies in the assembly's Hilbert space $\mathcal{H}$, as a sum of product states:
\begin{equation} \label{eq: psisimple}
|\psi\rangle = \sum_{i,j,k}c_{ijk}\ |\xi_i\rangle_1 \otimes |\eta_j\rangle_2 \otimes |\zeta_k\rangle_3\ ,
\end{equation}
where we have chosen a different single-particle basis for each particle (of course, there is no need to do this; we have done it merely to make the discussion more vivid).

A permutation $\pi$ on the particles reshuffles the particles amongst their single-particle factor states in each of the superposed product states in (\ref{eq: psisimple}).
For example, with two particles: the flip $\pi : 1\mapsto 2, 2\mapsto 1$, written as `(12)', defines the operator $P(12)$, whose action on product states is a flip: $|\xi_i\rangle_1\otimes|\eta_j\rangle_2 \in \mathfrak{h}_1\otimes \mathfrak{h}_2 \mapsto |\eta_j\rangle_1\otimes|\xi_i\rangle_2 \in \mathfrak{h}_1 \otimes \mathfrak{h}_2$, and which is extended by linearity to be a unitary operator.  In a similar way, we define a unitary operator on $\mathcal{H} :=  \bigotimes_l^n \mathfrak{h}_l$ representing an arbitrary permutation of $n$ objects.

It is harmless, and even useful, to allow a re-ordering of the factors, keeping track of them by the particle label subscripts; so that e.g.~we write equivalently  $P(12): |\xi_i\rangle_1\otimes|\eta_j\rangle_2 \in \mathfrak{h}_1\otimes \mathfrak{h}_2 \mapsto |\xi_i\rangle_2\otimes|\eta_j\rangle_1 \in \mathfrak{h}_1 \otimes \mathfrak{h}_2$.  This convention is useful for two reasons.  First, it permits a conveniently abbreviated way of writing down  entangled states, by using the distributive law for the tensor product.  Thus, for a state of three particles, in which particles 1 and 3 are entangled, we want to write
\begin{equation}
a|\xi\rangle_1 \otimes |\eta\rangle_2 \otimes |\zeta\rangle_3
 +
b|\zeta\rangle_1 \otimes |\eta\rangle_2 \otimes |\xi\rangle_3
\equiv
\Big(a|\xi\rangle_1 \otimes |\zeta\rangle_3
 +
b|\zeta\rangle_1  \otimes |\xi\rangle_3
\Big) \otimes |\eta\rangle_2 \ .
\end{equation}
This useful abbreviation was already available for entangled particles whose labels were adjacent numbers; we simply wish to extend this convenience to all pairs of entangled particles.  And, in fact, this is invaluable for explicit calculations with paraparticle states.  Second, this convention is useful for the analogy with Section \ref{subs:GR}'s presentation of a diffeomorphism and its relation to its pullback.

Returning to the case of three particles, a generic permutation $\pi$ on the particle labels defines a unitary operator $P(\pi)$.  Using our allowance to re-order the factors, we write:
\begin{equation}
P(\pi)|\psi\rangle = \sum_{i,j,k}c_{ijk}\ |\xi_i\rangle_{\pi(1)} \otimes |\eta_j\rangle_{\pi(2)} \otimes |\zeta_k\rangle_{\pi(3)}\ .
\end{equation}
For example, for the permutation $\pi: 1 \mapsto 2, 2 \mapsto 3, 3 \mapsto 1$, written `$(123)$', we have
\begin{equation}
P(123)|\psi\rangle = \sum_{i,j,k}c_{ijk}\ |\xi_i\rangle_2 \otimes |\eta_j\rangle_3 \otimes |\zeta_k\rangle_1\ .
\end{equation}

But of course, we could alternatively, and entirely equivalently, write $P(123)|\psi\rangle$ so that the particle labels preserve their order:
\begin{equation}
P(123)|\psi\rangle = \sum_{i,j,k}c_{ijk}\ |\zeta_k\rangle_1\otimes |\xi_i\rangle_2 \otimes |\eta_j\rangle_3
\end{equation}
But now it is clear that we could also have thought of $\pi$ as having induced a permutation not on the particles but on the single-particle factor \emph{states} in each superposed product state, so that particle $i$'s state is transferred to particle $\pi(i)$ under the permutation.  That is, particle $i$'s state \emph{alters}, under the permutation, to the state previously enjoyed by particle $\pi^{-1}(i)$.  We can think of this as a sort of ``pullback" permutation, on analogy with $d^*$ in Section \ref{subs:GR} (recall the definition $d^*g := g \circ d$).

We now return to the general case, in which $W$ may represent a mixed state.  Since $W$ may always be decomposed as a sum of projectors $W = \sum_i w_i |\psi_i\rangle\langle\psi_i|$, and the effect of the permutation $\pi$ on each vector $|\psi_i\rangle$ is to produce the new vector $P(\pi)|\psi_i\rangle$, $\pi$ induces a change of density operator from $W$ to $P(\pi)W P(\pi)^\dag$:
\\\\
\begin{tabular}{p{6cm}p{6cm}}
(\emph{Def $PW$}) &
$W \mapstochar\xlongrightarrow{\pi} P(\pi)WP(\pi)^\dag $ .
\end{tabular}
\\\\
\indent It is possible instead to consider $\pi$ as acting on the quantities.  In this case, it is best to define the action so that $Q$ goes to $P(\pi)^\dag QP(\pi)$ under the permutation.  It makes no difference to the expectation values whether we consider permutations to act on the states or the quantities: according to both conceptions, $Tr(W Q)$ goes to $Tr(P(\pi)W P(\pi)^\dag Q)$, because the trace operation is cyclic, i.e.~$Tr(P(\pi)W P(\pi)^\dag Q) \equiv Tr(W P(\pi)^\dag QP(\pi))$.  We emphasise that these two conceptions do \emph{not} align with our two construals of a permutation, described above. In fact, we will mostly consider (and it will be harmless to do so) the quantities and their eigenspaces to constitute a ``fixed background" in the Hilbert space, so that we conceive of a permutation as acting on the state $W$, as in (\emph{Def $PW$}).

\subsubsection{Symmetrisation and indistinguishability}\label{subsubs:SPIP}
It is common practice in QM to impose the \emph{symmetrisation postulate} (SP).  We present it here, as in our discussion of permutations, in two stages: first for rays and then for density operators in general.  SP for a ray $W = |\psi\rangle\langle\psi|$ is usually stated in terms of the vector $|\psi\rangle$: SP requires that $|\psi \rangle$ is either \emph{fully symmetr\textbf{ised}} in the sense of being fixed by all permutations,  $P(\pi)|\psi\rangle = |\psi\rangle$; or is \emph{fully antisymmetr\textbf{ised}} in the sense that for all $\pi$,  $P(\pi)|\psi\rangle = (-)^{\pi} |\psi\rangle$, where $(-)^{\pi}$ is the parity of $\pi$.
Note that these two sorts of vector form two orthogonal subspaces, ${\cal {S(H)}}$ and  ${\cal {A(H)}}$; they are usually called the {\em symmetric subspace} and the {\em antisymmetric subspace}.  The fully symmetrised states are known as \emph{bosonic}, and the fully antisymmetrised states are known as \emph{fermionic}.\footnote{Two remarks about jargon: (1) The textbook tradition varies as to whether the \emph{definition} of bosons and fermions involves their properties under odd permutations, as here, or instead their values of intrinsic spin, bosons having integer, and fermions half-integer, spin.  Famously, the two alternatives are linked by the \emph{spin-statistics theorem} (Pauli [1940]; see Massimi \& Redhead [2003] for a recent analysis); but this theorem belongs to quantum field theory, and so is outside our scope.
(2) More important for us: we will usually say `fully symmetrised', `fully symmetric', etc.~instead of the more common `symmetrised', `symmetric', etc.~because paraparticle states, which SP rules out, possess \emph{some} symmetries.\label{jargonremarks}}

 Let us say that $|\psi\rangle$ and its associated ray $|\psi\rangle\langle\psi|$ is \emph{fully symmetr\textbf{ic}} if and only if $|\psi\rangle$ is either fully symmetrised or fully antisymmetrised.  SP as applied to a \emph{ray} $W = |\psi\rangle\langle\psi|$ is therefore the requirement that $|\psi\rangle\langle\psi|$ be fully symmetric.  This requirement can be formulated neatly as follows:
 \\\\
 \begin{tabular}{rp{14cm}}
(\emph{SP-Ray}) & For all admissible states represented by a ray $|\psi\rangle\langle\psi|$ and every permutation $\pi$:
$P(\pi)|\psi\rangle\langle\psi|P(\pi)^\dag = |\psi\rangle\langle\psi|$.
\end{tabular}
 \\\\
 Thus stated, SP-Ray has two aspects, which we present in order of increasing importance for our purposes. (Agreed: for practical physics, and hence the textbooks, the order of importance is the reverse!)\\
\indent (i): As stated, SP-Ray forbids, as an admissible pure state, a linear superposition of bosonic and fermionic vectors. For $|\phi\rangle := c_1 |\psi_s \rangle + c_2 |\psi_a \rangle,$ with $|\psi_s \rangle \in \mathcal{S(H)}$  and $|\psi_a \rangle \in \mathcal{A(H)}$, and  $0 \neq c_i \in {\mathbb C}$, is neither fully symmetrised nor fully antisymmetrised. It is usual, and sometimes regarded as part of SP, to relax this veto and say instead that $|\phi\rangle$ represents a \emph{mixed} state with weights $| c_1 |^2$ and $| c_2 |^2$ of being in $|\psi_s \rangle$ and $|\psi_a \rangle$ respectively. (This of course makes sense for an ensemble, e.g.~a mixed beam of bosons and fermions.)
But we will instead take SP to insist that this mixed state be represented by the density operator $W_\phi :=  |c_1|^2 |\psi_s \rangle\langle \psi_s| + |c_2|^2 |\psi_a \rangle\langle \psi_a|$, which (see below) satisfies the generalisation of SP-Ray to cover all states.\\
\indent (ii):  It turns out that for two particles, $n = 2$, the fully symmetrised and fully antisymmetrised vectors together span ${\cal H} = \mathfrak{h}_1 \otimes \mathfrak{h}_2$: ${\cal {S(H)}} \oplus {\cal {A(H)}} = {\cal H}$. But for $n \geqslant 3$, ${\cal H}$ has {\em further} subspaces invariant under the action of all permutations, orthogonal to ${\cal {S(H)}} \oplus {\cal {A(H)}}$. These are the {\em paraparticle sectors}, which will be crucial to our revival of Stachel's argument in Section 5. For the moment, we just note that SP-Ray forbids states whose vectors lie in these subspaces.

We turn to formulating SP in terms of density operators. Recall that we may decompose (typically non-uniquely!) a density operator as a sum of weighted projectors:  $W = \sum_i w_i |\psi_i\rangle\langle\psi_i|$.  We extend our condition SP-Ray to any density operator $W$ by having SP require that $W$ have a decomposition in which all its projectors satisfy SP-Ray.  That is: a decomposition such that, for all $i$ and every $\pi$: $P(\pi)|\psi_i\rangle\langle\psi_i|P(\pi)^\dag = |\psi_i\rangle\langle\psi_i|$.  Note that this allows states which are an ensemble of bosons and fermions, in line with comment (i), above.

In the case of rays, we have the equivalence: $|\psi_i\rangle\langle\psi_i|$ satisfies SP iff $|\psi_i\rangle\langle\psi_i|$ commutes with every $P(\pi)$.  But in the case of density operators for three or more particles, this equivalence is weakened to a one-way implication. That is: since the property of commuting with every $P(\pi)$ is preserved under linear, in particular convex, combination, any density operator $W$ that satisfies SP is one for which, for every $\pi$, $P(\pi)WP(\pi)^\dag = W$.  But (as we will stress later in Section 5): for $n\geqslant 3$ (i.e.~allowing paraparticles), $W$ commuting with every $P(\pi)$ is \emph{not} sufficient for $W$ satisfying SP.  A trivial example is $W = \frac{1}{dim(\mathcal{H})}\mathbb{I}$, which for $n\geqslant 3$ does not satisfy SP, since any decomposition contains projectors into paraparticle sectors, as indicated in comment (ii).

Above we defined a ray to be fully symmetric iff its associated vector was fully (anti-)symmetrised.  But it is usual to define a \emph{symmetric} operator as one that commutes with all the $P(\pi)$s.  So in this jargon, we can sum up the preceding paragraph as: if $W$ obeys SP, then it is symmetric; but if $W$ is symmetric, it might not obey SP.  Being symmetric will be important, as a property of {\em quantities}, in our discussion of the indistinguishability postulate (IP); and again, as a property of either density operators or quantities, in Section 5.

As a summary, here is our general statement of SP in the form that will be most useful in our later discussion of Stachel's argument (cf.~Section \ref{subs:theories}):
\\\\
\begin{tabular}{rp{14cm}}
(\emph{SP}) & For all admissible states $W$ and every permutation $\pi$: there is a decomposition of $W = \sum_i w_i |\psi_i\rangle\langle\psi_i|$ such that for all $i$, $P(\pi)|\psi_i\rangle\langle\psi_i|P(\pi)^\dag = |\psi_i\rangle\langle\psi_i|$.
\end{tabular}
\\

A state's being bosonic or fermionic---its \emph{symmetry type}---leads to its statistics, i.e.~probabilities for various quantities, obeying certain constraints, which differ from those that are classically expected.  For example, no fermionic state may have repeating factors $|\phi\rangle_1\otimes|\phi\rangle_2\otimes\cdots|\phi\rangle_n$, since these states behave incorrectly under odd permutations (they do not pick up a minus sign); this is a case of \emph{Pauli exclusion}.  There are further consequences about statistics for both fermions and bosons; and some have (wrongly) argued that they imply that anti-haecceitism is true for these particles.  We will rehearse this argument in Section \ref{subsubs:badarg}, after considering the justification of SP.

There are many justifications for SP in the physics textbooks, of varying cogency.\footnote{A small selection: Schiff ([1968], p.~367), Gasiorowicz ([1974], pp.~147-8), Messiah ([1961], pp.~594-5) and Rae ([2006], pp.~205-6). Earman [2010] is a masterly discussion.} But the best justification is that it makes expectation values of physical quantities invariant under a particle permutation.  Recall that, in the state $W$, the expectation value of the quantity $Q$ is $Tr(W Q)$.  So SP is justified by its implementing the following requirement, the \emph{indistinguishability postulate} (IP):\footnote{Some may prefer to reserve `indistinguishability postulate' for the postulate which implements what we have called IP \emph{without} imposing SP.  This allows them to say, for example, that SP is a restriction on admissible \emph{states}, while `IP' is a restriction on admissible \emph{quantities}.  For us, it will be clearest to define IP so as to make it logically weaker than SP and uncontroversial (indeed compulsory!); so that SP and the restriction on quantities are alternative ways to implement it.  In this we follow French \& Redhead ([1988], p.~238) and French \& Krause ([2006], p.~134).}
\\\\
\begin{tabular}{rp{10cm}}
(\emph{IP}) & For all admissible quantities $Q$ and all admissible states $W$: $Tr(P(\pi)W P(\pi)^\dag Q) =Tr(W Q)$ for every permutation $\pi$.
\end{tabular}
\\\\
If $W$ obeys SP, then there is a decomposition $W = \sum_i w_i |\psi_i\rangle\langle\psi_i|$ for which, for every $i$ and $\pi$, $P(\pi)|\psi_i\rangle\langle\psi_i|P(\pi)^\dag = |\psi_i\rangle\langle\psi_i|$.  Therefore, for every $\pi$, $P(\pi)WP(\pi)^\dag = W$, and so IP is obeyed for \emph{any} $Q$.\footnote{A qualification on ``\emph{any} $Q$": We thank Nick Huggett for emphasising to us the point that, if SP is taken to restrict the space of kinematically possible states from $\mathcal{H}$ to either its symmetric and antisymmetric projections, $\mathcal{S}(\mathcal{H})$ or  $\mathcal{A}(\mathcal{H})$, then any admissible quantity $Q$ should, strictly speaking, be represented by an operator on this restricted space, rather than on $\mathcal{H}$.  In this case there is an effective restriction on admissible quantities, as well as states, by SP.  After such a restriction, the satisfaction of IP is somewhat over-determined: the admissible quantities are \emph{always} symmetric in this space, for the same reason that the admissible density operators are: viz.~that they are weighted sums of projectors that are all invariant under permutations (cf.~Section \ref{subsubs:511}).  We would like to point out, however, that, first of all, SP as we define it restricts the space of states to the direct sum $\mathcal{S}(\mathcal{H}) \oplus \mathcal{A}(\mathcal{H})$, which leaves plenty of room (in the off-diagonals between the boson and fermion states) for non-symmetric quantities.  Secondly, even if the space of states is restricted still further to either the boson or fermion subspace, one could alternatively take SP to be a restriction on initial states, so that non-symmetric (i.e.~paraparticle) states are still considered kinematically possible, though never in fact realised.  French \& Krause ([2006], p.~148) recommend this ``restricted intitial state" interpretation of \emph{IP} to the haecceitist.  \label{fn:Huggett}}

IP is compulsory: for it says that permuted states are experimentally, i.e.~statistically, equivalent, in that, for the expectation values of all physical quantities, it does not matter which particle plays which role.  This is compulsory because the particles are indistinguishable: they possess the same mass, charge and intrinsic spin, so that in determining the expectation values for physical quantities, it cannot matter which particle plays which role (cf.~Dieks ([1990], pp.~136-8), developed by Huggett ([1999], pp.~342-6)).\footnote{Huggett, in his work with Imbo ([2009]), has since given up on his claim that IP is compulsory.  There is no room here to fully represent the issues; but briefly, Huggett \& Imbo's rejection of IP seems to rely on 
thinking of the quantum formalism as ``opportunistic" about reference, i.e. saying that particle labels rely on non-overlapping single-particle spatial wavefunctions, which may be used as a basis for distinguishing the particles.   But such opportunism is not possible generally, and we would not want to rule out \emph{a priori} that favourable states in which it \emph{is} possible may evolve into unfavourable states, in which Huggett \& Imbo's strategy yields empirically incorrect results.  Therefore we retain IP \emph{pace} Huggett \& Imbo.}

SP is \emph{one} way of implementing IP; but there is another.  That is to restrict the class of admissible {quantities} \emph{without} restricting the admissible {states}.  Recall that, since traces are cyclic, $Tr(P(\pi)W P(\pi)^\dag Q) \equiv Tr(W P(\pi)^\dag Q P(\pi))$.  So we may satisfy IP, without imposing SP, by requiring that all admissible quantities $Q$ satisfy $P(\pi)^\dag Q P(\pi) = Q$, i.e.~$[Q, P(\pi)] = 0$, for all $\pi$.  Again, such quantities are called \emph{symmetric}.\footnote{To mesh with our characterization of states, we should, strictly speaking, call these quantities `fully symmetric': for there are also quantities which possess \emph{some}, but not all, symmetries.  However, we are only interested in the ``fully" symmetric quantities and we will only mention them; so we adopt the usual adjective, `symmetric', on its own.\label{jargonremarks2}}

Incidentally, it is worth noting that, even with SP in play, IP must still be imposed on a quantity of special importance, viz.~the Hamiltonian.\footnote{If we take SP not as a restriction on initial states, but as a restriction on kinematically \emph{possible} states, then all of the admissible quantities are symmetric anyway; cf.~footnote \ref{fn:Huggett}'s main discussion.}  For we surely require of our fully symmetric states that they preserve their symmetrisation under every physically possible evolution of the system.  That is, we demand that if $W$ is fully symmetric, then so is $U(t)W U(t)^\dag$ for every unitary $U(t)$ representing a physically possible evolution.  It is necessary and sufficient for this that $[U(t), P(\pi)] = 0$ for every $t$ and $\pi$.  Now, we can always write a unitary $U(t)$ as $e^{itH}$ for some self-adjoint $H$; this $H$ is by definition the Hamiltonian for the evolution.  And the commutation condition, $[U(t), P(\pi)] \equiv [e^{itH}, P(\pi)] = 0$, is ``inherited" by $H$: that is, $[H, P(\pi)] = 0$.  So every physically possible Hamiltonian is required to be symmetric, \emph{even when SP is imposed}.

Here we note an important qualification to our proposed analogy between GR and QM with regard to diffeomorphism/permutation invariance.  Recall (Section \ref{subs:GR}) that in GR, the diffeomorphism covariance of the dynamical equations allows one to construct two dynamically allowed histories which share a common past but may diverge thereafter. So the dynamical equations and boundary conditions' under-determination of mathematical states entails---as a special case---a certain form of indeterminism. (Originally, in 1913, Einstein saw this indeterminism as refuting diffeomorphism covariance. Earman and Norton [1987] use this to make the under-determination problem more vivid.)  It is important to understand that this is explained by two features which, combined, are peculiar to GR.  First, time is in some sense \emph{internal} to GR's physical states (remember that diffeomorphisms putatively shuffle around points of space\emph{time}). And second, the dynamical equations' symmetry group---viz.,~the diffeomorphisms---is incredibly rich, containing transformations that are the identity for certain large regions of the manifold, but diverge as wildly as you like elsewhere.  Since these features are peculiar to GR, the relevant generalisable result here is \emph{not} indeterminism, but rather: the under-determination of a unique mathematical solution by the dynamical equations and boundary conditions, subject to (perhaps a restricted form of) permutation symmetry. 

In the case of QM, we will \emph{not} find indeterminism---time just does not play an analogous role in QM---though, as we show, there is certainly under-determination.  The Hamiltonian, which defines the time-evolution of a quantum state, is always well-defined---as we have just seen, IP \emph{restricts}, rather than multiplies, the admissible Hamiltonians.  So, once we have \emph{identified} the system's state-vector (or density operator), its evolution is determinate.  Rather, the under-determination we wish to stress applies to the identification of the system's state-vector in the first place.  As we shall see, in the paraparticle sectors of a Hilbert space, a specification of the expectation values for \emph{all} of the admissible quantities under IP still fails to pick out a unique vector.\footnote{We should admit an important disanalogy between the models we have defined for GR and QM.  Histories, the models of GR, are specifications of field values on a four-dimensional manifold; so a history represents a possible world throughout all time.  On the other hand, density operators, the models of QM, specify expectation values only at a single time: they represent instantaneous states, not ``eternal" ones.  Besides, if in QM we shifted from the Schr\"odinger to the Heisenberg picture (cf.~footnote \ref{fn:SvsH}), the disanalogy would remain: cf.~also the last sentence of Section \ref{subsubs:perms}.}

We will return to IP later in Section \ref{s:para}; but for now let us take SP to be imposed. We are seeking an analogy to diffeomorphism covariance in GR, which will give anti-haecceitism about quantum particles some interpretative edge.  We now turn to what seems to us (and the literature!)~the \emph{only} salient candidate within the constraints of SP: an argument from quantum statistics.

\subsubsection{Quantum statistics and the bad argument for anti-haecceitism}\label{subsubs:badarg}
Recall that in GR histories are predicted by the dynamical equations not individually, but only up to diffeomorphism equivalence classes.  Under SP, quantum statistics can give the impression that, analogously, an individual ray in the Hilbert space does not represent a physical state; rather an isomorphism equivalence class of rays, connected by particle permutations, truly represents a physical state.  We shall first state this tempting argument; and then reject it.  The argument is (or variations on it are) widespread, even in the philosophy of physics literature.\footnote{E.g.~Schr\"odinger ([1984], p.~197), Reichenbach ([1956], pp.~66-71), and Teller ([1998], pp.~114-116).}  So far as we know, the fallacy of the argument was first exposed by French \& Redhead ([1988], pp.~235-8) and Dieks ([1990], pp.~133-4); our discussion will be a rehearsal of theirs.

Let us take the elementary and familiar example of two ``quantum coins"---systems each of whose Hilbert spaces are $\mathbb{C}^2$, so-called ``two-state" systems.  As we have seen, under SP these systems may be either one of two symmetry types, fermions or bosons.  Pauli exclusion for fermions means it will be most illuminating for us to consider \emph{bosonic} quantum coins, which obey \emph{Bose-Einstein} statistics.\footnote{A common example of a two-state system is a particle with intrinsic spin $\frac{1}{2}$.  In quantum field theory, the spin-statistics theorem requires that such particles be fermions.  We make no such requirement, because: (i) we work with fixed particle number, which is outside the scope of the theorem; and (ii) the Hilbert spaces need not represent states of \emph{spin}.}   We choose a basis in the single-coin Hilbert space, call it the ``heads-tails'' ($\{|H\rangle, |T\rangle\}$) basis; this induces a product basis, $\{|HT\rangle := |H\rangle_1 \otimes |T\rangle_2, \ldots \}$, in the two-coin assembly's Hilbert space.  Given an equal probability density over all rays in this space, and then conditionalizing on the particles being bosons, we obtain a probability of $\frac{1}{3}$ each for the states $|HH\rangle$ (two heads) and $|TT\rangle$ (two tails), and the remaining $\frac{1}{3}$ for the other two states, $|HT\rangle$ or $|TH\rangle$ (one head and one tail), \emph{combined}.  These contrast with the natural-seeming, classical Maxwell-Boltzmann probabilities of $\frac{1}{4}$ each.

Now we come to the slated analogy with diffeomorphism covariance. For there is a sense in which Bose-Einstein statistics seem ``not to care which is which" that, at first sight, is similar to the dynamical equations of GR failing to distinguish a history from its diffeomorphic cousins.  For if we group the basis states into permutation equivalence classes, then it is the Bose-Einstein statistics that appear the most natural, giving an equal probability of $\frac{1}{3}$ to each equivalence class.  Bosonic coins, it is therefore claimed, are best seen as anti-haecceitistic entities, in which the vectors $|HT\rangle$ and $|TH\rangle$ really describe the same physical state.

This argument is seductive---but wrong.  Later (Section \ref{subs:pooleyreply}) we will discuss more thoroughly why.  But the reader may already have noticed that the states $|HT\rangle$ and $|TH\rangle$ contravene SP.  This means that we originally picked out the wrong states as a basis: in fact, the state space for the bosonic coins is spanned by only \emph{three} orthogonal vectors, one of which contains $|HT\rangle$ and $|TH\rangle$ in equal superposition, namely $\frac{1}{\sqrt{2}}\left(|HT\rangle + |TH\rangle\right)$---and \emph{that} explains why the uniform probability measure assigns $\frac{1}{3}$ to each of the states.\footnote{More accurately: SP explains the quantum statistics \emph{in conjunction with the added fact} that the probability measure over the rays of Hilbert space is discrete rather than continuous; cf.~Saunders ([2006b], pp.~201-7).}  Therefore, unless SP already carries with it some prior commitment to structuralism (which it doesn't!---cf.~Section \ref{subs:SPah?}), quantum statistics offer no motivation for anti-haecceitism.\footnote{Of course, the statistics offer no motivation for haecceitism {either}.  We thank Nick Huggett for emphasising this to us.}

\section{The generalised hole argument for sets}\label{s:hasets}

We now shift focus from the particular to the general, by considering Stachel's `hole argument for sets', which is supposed to capture as instances the arguments for Leibniz equivalence in GR and anti-haecceitism in QM.  Stachel's presentation ([2002], pp.~236-9) contains some technical infelicities, which were detected and corrected by Pooley in his admirably careful assessment ([2006], pp.~104-114).  So throughout Section \ref{s:hasets}, we will stick to Pooley's reformulation.

We will first (Section \ref{subs:models}) outline the general idea of a model, belonging to a state space, as a mathematical representative of a possible world, i.e.~physical state; this general idea of a model is supposed to capture as instances both GR's histories and QM's density operators.  Then (Section \ref{subs:perms}) we will explicate the idea of an equivalence class of mutually isomorphic models all sharing the same domain.  In Section \ref{subs:syntax} we develop the same ideas in terms of syntactic, rather than semantic, concepts; this will also link structuralism to the Ramsey sentence approach and some ideas of Carnap [1950], viz.~state and structure descriptions.  In Section \ref{subs:theories} we give a conception of the structure of a physical theory, which is rough but adequate for our purposes; and which, developing the parallel between Section \ref{subs:perms} and Section \ref{subs:syntax}, comes in two versions, one semantic and one syntactic.  After all this setting up, Stachel's argument is given in Section \ref{subs:stachelarg}.

\subsection{Models and possible worlds}\label{subs:models}
 Our raw materials are a domain $\mathbf{S} = \{a_1,\ldots a_n\}$ of $n$ objects, each of which is given a name in the language, and an `ensemble' or `web' $\mathbf{R} = \{R_1, R_2,\ldots \}$ of $n$-place relations.  (For convenience, we imagine that each $m$-place predicate of the language where $m<n$ has been replaced by a suitably defined $n$-place predicate.\footnote{For finite domains, Stachel ([2002], p.~237) proposes the following device: for the $i$th $m$-place relation $Q_i(x_1, \ldots x_m)$ we define the corresponding $n$-place relation $R_i(x_1, \ldots x_n)$ by adding all possible relata in the remaining $n-m$ argument places.  That is, we define, for all $x_j$ ranging over $\mathbf{S}$:
$
\langle x_1, \dots x_m, x_{m+1}, \ldots x_n\rangle \in  \mbox{ext}(R_i)\ \mbox{iff}\ \langle x_1, \ldots x_m\rangle \in \mbox{ext}(Q_i).
$
We, like Stachel, consider this unproblematic; but it has been questioned (Jantzen [2009]).
\label{fn:rels}
})  We call the ordered pair $\langle \mathbf{S}, \mathbf{R}\rangle$ a \emph{model}.\footnote{Again, we use the term `model' without any connotation of satisfying some set of sentences; cf.~footnote \ref{fn:models}.}  It is a purely mathematical item, but it is intended to represent (perhaps partially) one or more physical states, which we will call, following Stachel and standard philosophy jargon, \emph{possible worlds}.  In GR, where the models are histories, we expect the domain $\mathbf{S}$ to be the four-dimensional manifold $M$, and the web $\mathbf{R}$ to encode the metric and matter fields $g$ and $T$.  We will not need details of this encoding, but it will build on Section \ref{subs:GR}'s treatment of the monadic case, i.e.~a relation between a point of $M$ and a value of the $g$-field (cf.~(\emph{Def $d^*g$})).\footnote{The fact that in GR, the domain $\mathbf{S} = M$ is uncountable seems to cause problems for defining the relations $R_i$ in the manner of footnote \ref{fn:rels}.  Following Stachel, and for the sake of an analogy with QM, we will imagine working with finite $n$.  This simplification will not undermine our main arguments and claims, in particular Section \ref{s:para}'s resurrection of Stachel's argument using paraparticles. Cf.~footnote \ref{fn:infinite2}. \label{fn:infinite1} }  In QM, some models are rays, some are convex combinations thereof;  the domain $\mathbf{S} = \{a_1, \ldots a_n\}$ is now the set of particles $\{p_1, p_2, \ldots p_n\}$, and the role of $\mathbf{R}$ is played by the density operator $W$.  Again, we do not need to specify the $R_i$ in detail, but we expect them to encode all unconditional and conditional Born rule single- and multi-particle probabilities, using the apparatus of partial tracing).  (It might have been expected that instead the \emph{quantities}, represented by self-adjoint operators, or perhaps their eigenspaces, would play the role of $\mathbf{R}$; but recall (Section \ref{subsubs:perms}) our convention that the quantities and their eigenspaces form a ``fixed background" in the Hilbert space.)

We also introduce the notion of a \emph{state-space} $\Sigma := \{\langle\mathbf{S}, \mathbf{R}\rangle\ |\ \mathbf{R}\ni R_i \subseteq \mathbf{S}^n \}$, which is the set of all models that share the same domain $\mathbf{S}$ and differ only as to the membership assignments of ordered $n$-tuples $\langle a_1, \ldots a_n\rangle \in \mathbf{S}^n$ to the relations $R_i \in \mathbf{R}$.  The idea is that the state-space represents the space of all possible worlds for the objects in $\mathbf{S}$, given the relations in $\mathbf{R}$---though for now we remain neutral as to whether this representation relation is one-to-one or many-to-one.  We also remain neutral as to whether $\Sigma$ contains \emph{every logically} possible assignment of objects in $\mathbf{S}$ to $\mathbf{R}$; that is, we allow the degree to which the space of possibilities is limited to vary from case to case.  But for the sake of definiteness, you can imagine that in GR, the state-space contains every kinematically possible history (that is, every suitably smooth assignment of field-values to all the spacetime points); and that in QM the state-space of pure states is given by (albeit with qualifications needed in Section \ref{s:para} for paraparticles) the rays in the Hilbert space $\mathcal{H} = \bigotimes_{l=1}^n \mathfrak{h}_l$.

\subsection{Permutations and permutes}\label{subs:perms}
Given this apparatus of models as representations of possible worlds, questions naturally arise concerning the representation relation:  Is it a surjection, i.e.~do the models leave anything out?  Is it one-to-many, i.e.~are the models in some way partial or incomplete representations? Is it many-to-one, i.e.~do the models contain redundant elements?

For our present discussion, we are concerned only with the last question, which concerns redundancy; and even then we only consider a special case of redundancy, whereby models differing by a permutation of indistinguishable objects represent the same possible world.  So let us ``suspend disbelief": that is, let us take our favourite theory (GR, QM, or some other ``fundamental theory'') to be complete, both in the sense that the representation is a surjection, and in the sense that each individual model is rich enough to represent a single possible world.  Then each possible world is represented by \emph{at least} one model in the state space.  Then our main question still remains: whether the representation relation is many-to-one; or, in other words, whether there is any redundancy in models' description of possible worlds; or, in yet other words, whether the models are trading in distinctions without a difference.

The redundancy we are concerned with relates to the roles of the objects in $\mathbf{S}$ in instantiating the web of relations $\mathbf{R}$.  For any model $\langle \mathbf{S}, \mathbf{R}\rangle$ we may construct the equivalence class of models which are isomorphic to it while sharing the same domain $\mathbf{S}$; call each such isomorphic copy a \emph{permute}, so that the equivalence class is the class of permutes.  Each permute is generated by one of the $n!$ permutations on the domain $\mathbf{S}$.  Writing a permutation on $\mathbf{S}$ as $\mathbf{P}$ (the generalisation of the diffeomorphism $d$ in GR, and $\pi$ in QM), and writing `\emph{s}' as a shorthand for $\langle s_1, s_2, \ldots s_n \rangle$, a possibly repeating ordered $n$-tuple of objects in $\mathbf{S}$, so that each $s_i \in \{a_1, a_2, \ldots a_n\}$, we write `$\mathbf{P}(s)$' for the sequence arising from applying $\mathbf{P}$ to $s$.  That is: $\mathbf{P}(s) := \langle \mathbf{P}(s_1),  \mathbf{P}(s_2), \ldots  \mathbf{P}(s_n) \rangle$. $\mathbf{P}$ also induces an ``isomorphic copy" $\mathbf{PR} := \{\mathbf{P}R_1, \mathbf{P}R_2,\ldots\}$ of the web of relations $\mathbf{R}$, according to the definition (cf.~Pooley [2006], p.~105):
$$
(\mbox{\emph{Def PR}}) \qquad\qquad \forall s \in \mathbf{S}^n,\ \forall R_i\in \mathbf{R}: \ \ s \in \mathbf{P}R_i \ \ \mbox{iff}\ \ \mathbf{P}^{-1}(s) \in R_i\ .\qquad\qquad
$$
(Recall (\emph{Def $d^*g$}) in Section \ref{subs:GR}, and (\emph{Def $PW$}) in Section \ref{subsubs:perms}.)

For some models, non-trivial (i.e.~non-identity) permutations do not always generate a new permute.  These models are of particular interest.  Let us first define a \emph{symmetry} of a model $\langle \mathbf{S}, \mathbf{R}\rangle$  as a permutation $\mathbf{P}$ on the objects in $\mathbf{S}$ that returns the original model:  i.e.~a $\mathbf{P}$ for which $\langle \mathbf{S}, \mathbf{PR}\rangle$ = $\langle \mathbf{S}, \mathbf{R}\rangle$.  A \emph{symmetric model} is then defined as a model which possesses at least one non-trivial (i.e.~non-identity) symmetry.   Amongst the symmetric models are the \emph{fully symmetric} models, each of which only has one permute, namely itself; for them \emph{every} permutation is a symmetry.  In summary, we define:
\begin{center}
\begin{tabular}{lccc}
$\langle \mathbf{S}, \mathbf{R}\rangle$ is \emph{symmetric}
& iff &
for some $\mathbf{P} \neq $\ \textit{identity}, & $\langle \mathbf{S}, \mathbf{PR}\rangle = \langle \mathbf{S}, \mathbf{R}\rangle$;\\
$\langle \mathbf{S}, \mathbf{R}\rangle$ is \emph{fully symmetric}
& iff &
for every $\mathbf{P}$, & $\langle \mathbf{S}, \mathbf{PR}\rangle = \langle \mathbf{S}, \mathbf{R}\rangle$.
\end{tabular}
\end{center}
Note that this general definition of a fully symmetric model includes as an instance our definition of a fully symmetric ray in QM, as required by SP-Ray (Section \ref{subsubs:SPIP}).  This is our first hint of the strength, and so doubtfulness, of SP.

\subsection{State descriptions and structure descriptions}\label{subs:syntax}
So far, we have presented models not in the language of the theory in question (GR or QM), but in the informal metalanguage.  Models may also be specified in the language of the theory itself; but so long as the domain is finite (where we can achieve categoricity), and so long as we imagine that the primitive vocabulary of the theory is endowed with an intended interpretation, the difference is one only of taste.\footnote{For infinite domains, we must contend with the L\"owenheim-Skolem theorem, but even then there we could resort to infinitary languages.  In these languages, we can define relations between infinitely many objects; cf.~footnote \ref{fn:infinite1}. \label{fn:infinite2}}

But it is worth describing how models, and properties like being symmetric, are presented in the theory's language.  For Pooley uses these ideas---especially the ideas of state descriptions and structure descriptions---to overcome the technical problems in Stachel's account.  So we join Pooley in this, because we want to be faithful to his discussion, and also to keep in mind the parallels between structuralism as we have defined it, and `Ramsey-sentence realism' (mentioned in Section \ref{subs:structure}).

Such a specification of a model may be called (following Carnap [1950], p. 71) a \emph{state description}.  It is a sentence formed by conjoining instances of the following three types of formula:
\begin{enumerate}
		\item For each (possibly repeating) $n$-tuple $s  \in \mathbf{S}^n$ of objects in the domain, and each relation $R_i \in \mathbf{R}$, we have either the formula $R_i(s)$ or the formula $\neg R_i(s)$, depending on whether or not $s \in$ ext$(R_i)$ according to the model $\langle \mathbf{S}, \mathbf{R} \rangle$.\footnote{We use `$R_i$' to represent not only the $i$th relation in $\mathbf{R}$, but also the \emph{predicate} in the primitive vocabulary, which, under the intended interpretation, is \emph{assigned} the relation $R_i$.  Likewise, `$a_i$' denotes both an object in $\mathbf{S}$ and the name which, under the intended interpretation, is assigned that very object.}
		\item For each pair of distinct objects $a_1, a_2 \in \mathbf{S}$ we have the formula $a_1 \neq a_2$.  The conjunction of all these formulas imposes the restriction: \emph{there are at least $n$ objects}.
		\item There is only one type 3 formula.  It is a disjunction within the scope of a universal quantifier: $\forall y (\bigvee^n_{i=1} y = a_i)$.  It says: \emph{there are at most $n$ objects}.\footnote{Formulas of type 2 and 3 are either superfluous or potentially contradictory if one adopts a structuralism that lies at the strong end of the spectrum described in Section \ref{subs:structure}.  For these structuralisms, identity facts are already accounted for by the type 1 formulas. These matters are further discussed in Caulton \& Butterfield [2010a].}
\end{enumerate}
Following Pooley, we write the conjunction of all these formulas as $\mathcal{R}(a)$, or, to make the argument places explicit, $\mathcal{R}(a_1, \ldots a_n)$, where $a_1, \ldots a_n$ is a \emph{non-repeating} sequence of \emph{names} for all the objects in $\mathbf{S}$.  In the intended interpretation, the sentence $\mathcal{R}(a)$ is true only of the model $\langle\mathbf{S}, \mathbf{R}\rangle$.

Using state descriptions, we can syntactically capture the idea of a permutation of objects embedded in the web of relations.  Accordingly, we can provide a syntactic counterpart of the idea of a permute, and also syntactic counterparts of our previous definitions of symmetric and fully symmetric models.  Given the definition (\emph{Def PR}) in the previous Section, it is clear that, just as  the sentence $\mathcal{R}(a)$ is true (under the intended interpretation) only of the model $\langle\mathbf{S}, \mathbf{R}\rangle$, so too the sentence $\mathcal{R}\left(\mathbf{P}^{-1}(a)\right)$ is true  (under the intended interpretation) only of the model $\langle\mathbf{S}, \mathbf{PR}\rangle$.   Therefore, quite generally we may call $\mathcal{R}(a)$ and $\mathcal{R}\left(\mathbf{P}(a)\right)$ \emph{permutes} of one another. We also extend the notions of being symmetric and being fully symmetric to state descriptions, in the obvious way.  We define:
\begin{center}
\begin{tabular}{lccc}
$\mathcal{R}(a)$ is \emph{symmetric}
& iff &
for some $\mathbf{P} \neq$ \ \textit{identity}, & $\mathcal{R}(a) \vdash \mathcal{R}\left(\mathbf{P}(a)\right)$;\\
$\mathcal{R}(a)$ is \emph{fully symmetric}
& iff &
for every $\mathbf{P}$, & $\mathcal{R}(a) \vdash \mathcal{R}\left(\mathbf{P}(a)\right)$.
\end{tabular}
\end{center}

The state description $\mathcal{R}(a)$ implies a logically weaker sentence, which we call (again, following Carnap) a \emph{structure description}.  The idea is that a structure description expresses the content of a corresponding state description \emph{except} for the information \emph{how}---i.e.~in what order---the objects in $\mathbf{S}$ instantiate the relations in $\mathbf{R}$.  To obtain the structure description implied by a given state description, we replace each occurrence of a name $a_i$ in $\mathcal{R}(a_1, \ldots a_n)$ with an instance of the free variable $x_i$, and do this for all $i=1, \ldots n$, and then existentially quantify over all the $x_i$s to obtain $\exists x_1 \cdots \exists x_n\mathcal{R}(x_1, \ldots x_n)$.  This sort of sentence in general picks out, not a single model $\langle\mathbf{S}, \mathbf{R}\rangle$, but the entire equivalence class of its permutes, $\{\langle\mathbf{S}, \mathbf{PR}\rangle\ |\ \mathbf{P}: \mathbf{S} \leftrightarrow \mathbf{S}\}$.\footnote
{Strictly speaking, the existentially quantified sentence picks out \emph{any} model with the structure description's pattern of instantiation $\mathcal{R}(x_1,\ldots x_n)$, \emph{even models with objects other than those in} $\mathbf{S}$.  Agreed, if we only allow extensional interpretations of the predicates in $\mathcal{R}$, then trivially this is ruled out, since then we interpret the predicates in  $\mathcal{R}$ by specifying a $\mathbf{R}$, which contains the objects in $\mathbf{S}$.  But if we allow \emph{intensional} interpretations of the predicates in $\mathcal{R}$, the interpreted structure description is not just blind as to `which is which'; it is also blind as to what the domain actually contains.  In any case, this double blindness will not affect our arguments.\label{fn:blind}}
So a structure description succeeds in picking out a unique model when and only when that model is fully symmetric; for then the equivalence class of permutes is a singleton set.

As an incidental remark, we note that the structure description, which the structuralist takes as a complete specification of a possible world, clearly has deep similarities to the Ramsey sentence of a theory, mentioned in Section \ref{subs:structure}.  For both Ramsey sentences and structure descriptions, the objects discussed (i.e.~quantified over) are conceived ``functionally", i.e.~purely in terms of which role they each play in the structure.  And for both, this conception has been celebrated as a means to eliminate a spurious under-determination (by e.g.~French \& Ladyman [2003]).

\subsection{Theories; permutability, fixity and general permutability}\label{subs:theories}

Stachel's argument requires some general conception of a theory, which captures GR and QM as examples.  It is of course difficult and controversial to give a detailed and plausible account of physical theories in general.  We will propose two ``caricatures", one syntactic and one semantic. Though they are evidently inadequate as general conceptions of a theory, they \emph{are} adequate for Stachel's, and our, purposes.

Syntactic caricature first: We may define a \emph{theory} in the traditional sense of logic, i.e.~as a set of sentences closed under logical deduction.  We can even be more liberal: any set $T$ of sentences may be a theory.  We add to this, the idea of a set $\Theta \ni \theta$, where $\theta$ is a statement of facts that are used, in conjunction with a theory's laws, to make predictions.  For short, we shall call such a $\theta$ \emph{initial or boundary conditions.}  So each $\theta$ is a sentence---perhaps a horrendously long conjunction---which describes a contingent fact on the basis of which $T$ is supposed to make predictions.  In GR, a $\theta$ will be a statement about field values for regions of spacetime, most notably a specification of exact values throughout a spacelike hypersurface; in QM, a $\theta$ will be a specification of expectation values for one or more admissible quantities at a time, most notably a specification of eigenvalues of a complete commuting set of quantities.  A typical $\theta$ will not suffice to yield a unique prediction, given the sentences in $T$.  But there may be \emph{some} $\theta$s---we might call them \emph{maximal} initial or boundary conditions---which do yield a unique prediction.  If the subset of maximal $\theta$ is suitably rich (e.g.~if for each time about which we wish to predict, there is a maximal $\theta$ for each earlier time), then we will say that $T$ is a deterministic theory.

But the converse does \emph{not} hold: that is, a theory may yet be deterministic \emph{even if} it fails to yield a unique prediction given initial or boundary conditions that are as rich as possible.  For, as we have stressed, we may overcome the under-determination in an interpretative move, viz.~by taking the predicted models to represent many-to-one a physically possible world.  This is precisely what structuralism aims to do: restore determinism by collapsing the apparent distinction between permutes.

Now the semantic caricature: Alternatively, a theory $T$ may be defined as an ordered pair $T = \langle\Sigma, \sigma \rangle$, where $\Sigma$ is a state-space (in the sense of Section \ref{subs:models}) and $\sigma$ is a \emph{selection function} (which we might also call a \emph{prediction function}).  The selection function $\sigma: \Theta \to \mathfrak{P}(\Sigma)$ = the power-set of $\Sigma$, takes as argument some initial or boundary conditions $\theta \in \Theta$ and returns as value some subset $\sigma(\theta) \subseteq \Sigma$ of the state-space.  Here we may take $\Theta$ as before, as a set of sentences; or we may take $\Theta = \mathfrak{P}(\Sigma)$, where each $\theta \in \Theta$ is the set of models in which a certain collection of initial or boundary conditions hold true: nothing hangs on the decision, so we shall leave it undecided.  The idea is that $\sigma$ plays an analogous role in the semantic conception as sets of sentences, together with logical entailment, play in the syntactic conception.  So, similarly to the syntactic conception, we may worry that $T$ is indeterministic if, even for $\theta$ as rich and detailed as possible, the value $\sigma(\theta)$ of $\sigma$ is typically not a singleton set.  And again, similarly, a structuralist aims to calm the worry, and restore determinism, for the case where $\sigma(\theta)$ is a set of permutes.

With all this set up, we are now in a position to discuss two formal features that a theory, or its selection function, may exhibit, and which will be of crucial importance in the following.  These are \emph{permutability}, which Stachel ([2002], p.~238) and Pooley ([2006], p.~106) both discuss, and \emph{fixity}, which we wish especially to articulate and discuss.\footnote{The notion of fixity is present in Pooley ([2006], p.~113, parag.~1), where he distinguishes between symmetry of the theory (i.e.~permutability) and symmetry of a solution of the theory (which is required for fixity).}  We now present these, giving equal attention to the syntactic and semantic conceptions.

\begin{description}
\item[Permutability:] Following Pooley (who extends Stachel's jargon to apply to theories), let us say that a theory is \emph{permutable} if and only if it selects models, or deductively entails state descriptions, invariantly over their equivalence classes of permutes.  That is:

(\emph{Sem})\ $T$ is permutable iff for all $\theta$, for every $\mathbf{P}$: $\langle\mathbf{S}, \mathbf{R}\rangle \in \sigma(\theta) \Leftrightarrow \langle\mathbf{S}, \mathbf{PR}\rangle \in \sigma(\theta)$;

(\emph{Syn})\ $T$ is permutable iff for all $\theta$, for every $\mathbf{P}$: $T, \theta \vdash \mathcal{R}(a) \Leftrightarrow T, \theta \vdash \mathcal{R}(\mathbf{P}(a))$.

With one qualification, a good example of permutability is our ``old friend", diffeomorphism covariance in GR.  The qualification is that diffeomorphisms constitute a proper \emph{subgroup} of the group of all permutations of manifold points.  We can accommodate this qualification by relativising permutability: we may say that GR is a theory which is \emph{permutable under smooth permutations with a smooth inverse.}

An everyday (and so more rough!), example of permutability is given by a team of renovators, each one a Jack of all trades.  We may have one renovator responsible for the electrical wiring, one for the plumbing and a couple for the painting and decorating.  If we permute the renovators, then we clearly change the profile of the team (we change who does what); but thanks to their versatility, we can predict that the house is equally well restored.

\item [Fixity:] We will also say that a theory is \emph{fixed} or has \emph{fixity} if and only if it selects \emph{only} models, or deductively entails \emph{only} state descriptions, that are \emph{fully} symmetric.  That is:

(\emph{Sem})\ $T$ has fixity iff for all $\theta$: $\langle\mathbf{S}, \mathbf{R}\rangle \in \sigma(\theta) \ \Rightarrow \ $ for every $\mathbf{P}, \ \langle\mathbf{S}, \mathbf{PR}\rangle = \langle\mathbf{S}, \mathbf{R}\rangle$;

(\emph{Syn})\ $T$ has fixity iff for all $\theta$: $T, \theta \vdash \mathcal{R}(a) \ \Rightarrow\ $ for every $\mathbf{P},\ \mathcal{R}(a)\vdash\mathcal{R}(\mathbf{P}(a))$.

An example of fixity is provided by SP in QM (cf.~Section \ref{subsubs:SPIP}), in which the objects are the particles $p_1, p_2, \ldots p_n$.  Recall that SP implies that any admissible state be given by a density operator with a decomposition all of whose rays are invariant under every particle permutation: $|\psi_i\rangle\langle\psi_i| = P(\pi)|\psi_i\rangle\langle\psi_i|P(\pi)^\dag$.  In Section \ref{s:fixed}, this example will be crucial to the disanalogy between GR and QM that Pooley emphasises.

A rough example (that is less quotidian than the renovators!)~is provided by a group of monastic scribes, each reproducing the same text.  If we permute the scribes, then clearly no difference results, for each was performing the same role anyway.
\end{description}

It is trivial that fixity entails permutability, on both the syntactic and semantic construals (\emph{Syn}) and (\emph{Sem}).  But a theory may be permutable and unfixed; GR is such an example.

Permutability and fixity link in with two other terms, familiar in the philosophy of spacetime physics.  These are \emph{covariance} and \emph{invariance}.\footnote{Cf.~Brading \& Castellani ([2006], p.~1343), citing Ohanian \& Ruffini [1994].}  To take a familiar example: in special relativity,  a Poincar\' e transformation is a special kind of permutation of the spacetime points (more special still than a diffeomorphism).  Permutability under Poincar\' e transformations is then nothing but (active) Lorentz covariance.  Fixity under Poincar\' e transformations is a far stricter requirement: the contents of Minkowski spacetime must be \emph{invariant} under arbitrary translations, rotations and boosts.

One could be forgiven for thinking that the fixity condition is so strong as to destroy all prospects for heterogeneity, and thereby that fixity cannot be obeyed by any remotely empirically successful theory.  However, this is not so for QM with SP, because of the superposition principle.  This allows us to superpose heterogeneous states appropriately, so that the universe may remain ``interesting", while \emph{obeying} fixity: roughly speaking, in these states, each particle plays an equal role in producing the heterogeneity.  As a result, every particle of the same species occupies the same (mixed) state.\footnote{As mentioned in footnote \ref{fn:HB} and following: this result has been used by various authors, e.g.~Margenau [1944], French \& Redhead [1988], Butterfield [1993] and Huggett [2003] to argue for the failure in QM of a certain formulation of Leibniz's principle of the identity of indiscernibles: an argument that has recently been questioned by Muller \& Saunders [2008] and Muller \& Seevinck [2009].  But it is not emphasised enough how very counter-intuitive this result is: Pooley is an exception ([2006], p.~116).}

We emphasise that permutability and fixity are \emph{formal} features that a (precise enough formulation of a) theory may have or lack; they do not depend on any interpretative stance one may take on the model-to-possible-world representation relation.  One such stance---viz.~claiming a many-to-one relation between models and possible worlds---is relevant to us. Once again we borrow the idea, and the name, from Stachel ([2002], p.~242) and Pooley ([2006], p.~106):
\begin{description}
\item[General permutability:] A theory is taken to be \emph{generally permutable} if and only if it is interpreted as treating as equivalent representations that differ only about which object plays which role in the instantiation of the web of relations.  To be more precise:

(\emph{Sem}) $T$ is generally permutable iff permutes are taken to represent the \emph{same} possible world; that is, iff the entire equivalence class $\{ \langle \mathbf{S}, \mathbf{PR}\rangle \ |\ \mathbf{P}: \mathbf{S}\leftrightarrow\mathbf{S}\}$ is taken to represent a single possible world.

(\emph{Syn}) $T$ is generally permutable iff the structure description $\exists x_1\cdots\exists x_n\mathcal{R}(x_1, \ldots x_n)$ (perhaps supplemented with a specification of a domain $\mathbf{S}$; cf.~footnote \ref{fn:blind}) is taken to describe a single possible world.

Thus general permutability captures, as instances, Leibniz equivalence in GR and anti-haecceitism in QM.  We take general permutability to be the formal statement of our notion of structuralism, which is to say that it captures all and only the structuralist positions lying on Section \ref{subs:structure}'s spectrum.  Recall that all those positions take individuality of objects to be grounded in their qualitative properties and relations.  This claim \emph{alone} motivates the ascent from models to their equivalence class of permutes, as the representation of a possible world. The remaining issue of disagreement amongst the various positions on the spectrum, viz.~whether diversity too is qualitatively grounded or else primitive, has no bearing on this indifference as to ``which is which".

\end{description}

Now, while general permutability, in contrast to permutability and fixity, \emph{is} an interpretative stance, it is subject to constraints, which arise from those two formal features.  Firstly, a theory must be permutable for it to be generally permutable.  Otherwise, for some $\theta$, it would predict some permutes and not others: which would be to recognise a physical difference between supposedly equivalent representations.  Secondly, a theory must be generally permutable if it has fixity. For in this case there is not a multiplicity of representations for which to claim equivalence: each set of models selected by the theory is a singleton, containing only one, fully symmetric model.  So one's freedom to adopt general permutability for a theory is constrained on both sides:  permutability is a necessary condition; fixity is sufficient.  Let us call this the \emph{general permutability constraint} ({GPC}):

\begin{tabular}{rl}
(\emph{GPC}) & $T$ has fixity $\ \Rightarrow\ $ $T$ is generally permutable $\ \Rightarrow\ $ $T$ is permutable.
\end{tabular}

We have already seen in the specific cases of GR and QM (Sections \ref{subs:GR}-\ref{subs:QM}) the key attraction of general permutability: namely, that it purports to explain away, or make palatable, the under-determination (in particular, indeterminism) arising from permutability.  Given the most complete possible initial or boundary conditions, a theory may fail to pick out a unique model from amongst its permutes.  But we may deny physical under-determination by invoking general permutability.  This is the core of Stachel's argument.

\subsection{Stachel's argument for structuralism}\label{subs:stachelarg}

In Section \ref{subs:structure}, we defined structuralism as the position that the individuality, and perhaps also the numerical diversity, of objects is grounded in their pattern of instantiation of qualitative properties and relations.  It should be clear, therefore, that structuralism about a theory $T$ implies a commitment to $T$'s being generally permutable.  Stachel goes further ([2002], p.~242): for him structuralism about $T$ is \emph{equivalent} to holding $T$ to be generally permutable.  Stachel uses this, together with general permutability's interpretative advantage in the case of permutable theories, to launch his abductive argument for structuralism about those theories.

However, Stachel does not present his argument in a premises-and-conclusion form: our presentation has been a reconstruction based on his discussion. So for each line of our reconstruction, we give the appropriate supporting quotations from his paper.

Given a permutable theory $T$:
\begin{enumerate}
	\item Structuralism about  $T$ is defined by holding general permutability for $T$.\footnote{
	`\ldots in a generally permutable world, insofar as the ensemble of relations $\mathbf{R}$ is concerned, the individuation of an entity such as $a_1$ in that world depends \emph{entirely} on the place it occupies in the ensemble of relations $\mathbf{R}$'  (p.~242, parag.~2).  Compare our definition of structuralism in Section \ref{subs:structure}.
	}
	\item General permutability for $T$ explains/makes palatable the permutability of $T$.\footnote{
	(Slight corrections have been made, in light of Pooley's criticism:) `\ldots if the entities are generally permutable, \ldots then $\mathbf{R}(a)$ [correction: $\langle\mathbf{S},\mathbf{R}\rangle$ or $\mathcal{R}(a)$] and $\mathbf{R}(Pa)$ [correction: $\langle\mathbf{S},\mathbf{P^{-1}R}\rangle$ or $\mathcal{R}(\mathbf{P}(a))$] are the same state for every permutation [$\mathbf{P}$]. \ldots This is the generally permutable escape from the hole argument' (p.~245, parag.~2).  Or, in the specific instance of GR: `\ldots if we are not to abandon the Einstein equations---or indeed \emph{any} covariant set of field equations---there must be a way to evade the conclusion of the hole argument.  That way is clear: it is to deny that diffeomorphically related mathematical solutions to the field equations represent physically distinct solutions' (p.~233, parag.~4).
	}
	\item Therefore (by inference to the best explanation): the permutability of $T$ supports structuralism about $T$.\footnote{Cf.~p.~236, parag.~6.}
\end{enumerate}

This argument certainly captures the standard argument for Leibniz equivalence in GR.  However---as Pooley was first to see---it seems to lose its grip over QM, when SP is assumed; for SP yields fixity, not just permutability.  But this problem is not specific to QM; it holds generally for any fixed theory. In fact, the topic of fixity brings both premises into question, as we shall now see.

\section{Fixed theories and metaphysical under-determination}\label{s:fixed}

\subsection{Pooley's objection; amending Stachel's premises} \label{subs:pooleyreply}

Pooley's objection ([2006], p.~109f.)~may be reconstructed in two parts, each corresponding to one of the premises of our reconstruction of Stachel's argument.  In this section, we will heed Pooley's objections and suggest revised premises by which Stachel's argument will be revived (in Section \ref{s:para}) for QM with paraparticles.

Let us take premise 1 first.  This is the claim that structuralism about a theory $T$ is equivalent to holding general permutability for $T$.  Now it is certainly true, given the discussion in Section \ref{subs:structure}, that structuralism about $T$ is \emph{sufficient} for holding $T$ to be generally permutable.  But a look at GPC shows that structuralism is not necessary for general permutability.  If a theory is fixed, then trivially it is generally permutable.  But fixity is a \emph{formal} feature of a theory---it applies to a theory or not, prior to any interpretative stance.  So no matter what our metaphysical position about the identity and individuality of a theory's objects, we are committed to its being generally permutable, if it is fixed.  Structuralism about a theory cannot therefore be \emph{defined} as holding general permutability for that theory: for structuralism is a substantive metaphysical thesis, yet we may be automatically committed to general permutability of a theory, by its formal structure.

So we should amend Stachel's definition of structuralism to exclude the case of fixed theories.  This is done easily enough: we say instead that

$1'$. Structuralism about \emph{unfixed} $T$ is defined by holding general permutability for $T$.

\noindent Here we recall Section \ref{subs:structure} and Section \ref{subs:theories}'s definition of general permutability, and reiterate that we intend $1'$ as a definition of structuralism which captures Section \ref{subs:structure}'s full spectrum of positions.  We admit that in the case of a fixed theory it makes \emph{no difference} whether one is a haecceitist or a structuralist: either way, the theory is general permutable.  Accordingly, holding a fixed theory to be generally permutable is an ``empty gesture": it does not commit one to a metaphysical position about the identity or individuality of the theory's objects.

Now we address premise 2. This is the claim that general permutability explains, or makes palatable, permutability.  But this claim falsely presumes that permutability always produces some sort of puzzle, or quandary, in need of explanation.  We characterised permutability by saying that a permutable theory selects all the permutes of any selected model.  But this is only a case of under-determination---and so some sort of puzzle---when the equivalence class of permutes contains more than one model.  This is not so for a fixed theory.  In this case, the theory is permutable but unproblematically so: it is an immediate consequence of fixity.

Accordingly, we adjust premise 2 in the obvious way:

$2'$. General permutability for \emph{unfixed} $T$ explains/makes palatable the permutability of $T$.

\noindent In the case of fixed $T$, permutability creates \emph{no} puzzle: permutability is ``explained" (if such a trivial result deserves that description) by the theory's fixity.  Just as well: for in the case of fixed $T$, general permutability, which by $2'$ explains permutability in the case of unfixed $T$, is as unavoidable a fact as permutability.

To sum up: a fixed theory picks out unique models (assuming it is not otherwise indeterministic)---and so there is no general interpretative advantage, or disadvantage, of \emph{either} haecceitist or structuralist interpretation over the other. In the familiar philosophical jargon: when a theory is fixed, its metaphysics of identity and individuality is \emph{under-determined}.\footnote{As Pooley ([2006], p.~111, parag.~1) observes: `If $\mathcal{R}(a)$ is a description of a world allowed by quantum mechanics, then $\mathcal{R}(Pa)$ will be a description of \emph{exactly the same world} (it will be a logically equivalent formula).  Thus stipulating that $\exists x_1\ldots\exists x_{(n-m)}\mathcal{R}(b, x_1, \ldots x_{(n-m)})$ \emph{does} suffice to pick out a unique world, \emph{even if one believes that the individuality of quantum particles transcends the relational structure in which they are embedded}' (his emphasis).}

Consider the specific example of QM with SP, and recall the fallacious argument for anti-haecceitism using bosonic coins (Section \ref{subsubs:badarg}).  In fact it is SP---and not anti-haecceitism---that explains the quantum statistics.  For, once SP is imposed, the Hilbert space for the two bosonic coins is spanned by only three states, all of which are fully symmetrised.  And a uniform measure over these rays yields the Bose-Einstein probabilities of $\frac{1}{3}$ for each outcome.  Nothing is gained by ascending to equivalence classes of permutes, for they are singletons: we obtain the same three states, each of them just surrounded by set-theoretic brackets.

This specific claim of under-determination, made by Pooley and endorsed by us, harks back to the repeated insistence by French\footnote{Cf. footnote \ref{manyvoices}.} that the physics of quantum statistics under-determines the metaphysics of individuality for quantum particles.   Pooley's objection to Stachel brings French's point into relief. It is not just an example of the familiar and general point that metaphysics is under-determined by physics:  viz.~that we can almost always contrive to save our favourite metaphysical doctrine in the face of a physical theory.  The point is much more specific: that, thanks to fixity, haecceitism and anti-haecceitism are \emph{equally natural} metaphysical doctrines in the context of QM with SP.

\subsection{Is SP anti-haecceitistic?}\label{subs:SPah?}
Here is the place to refute a tempting but mistaken view that, regardless of our foregoing discussion, anti-haecceitism \emph{is} supported by QM with SP, \emph{since anti-haecceitism is the original motivation for SP}.  If this were true, then it would be hardly surprising that anti-haecceitism fails to do any explanatory work \emph{after} SP has been accepted.

However, anti-haecceitism is neither necessary nor sufficient as a motivation for SP. It is \emph{not necessary} for SP because even a haecceitist has a motivation for SP: namely that it is a natural and expedient way to implement IP (cf.~Section \ref{subsubs:SPIP}).  Recall that IP is compulsory, for haecceitists as well as anti-haecceitists, since it merely expresses the \emph{physical indistinguishability} of permuted states, not their identity.  So whatever our metaphysical stance, we must impose IP.  SP is then simply one way of implementing IP; a way that is calculationally convenient and, given the apparent non-existence of paraparticles, harmless.

Nor is anti-haecceitism \emph{sufficient} for SP.  It is true that, under SP, each ray, indeed each density operator, is invariant under a particle permutation.  And anti-haecceitism demands that a physical state is invariant under a particle permutation.  \emph{But this implies SP only if we demand that physical states are represented by single rays.}  If we hold (as allowed by general permutability) a many-to-one representation relation between models (density operators) and worlds, then we can retain anti-haecceitism without requiring our models to be fully symmetric.  It is precisely this consideration that leads us to consider QM \emph{without} SP.  We shall see in the following Section that, by relaxing SP, we may remove QM's fixity and thereby bring it within the scope of the modified premises $1'$ and $2'$ of Stachel's argument for structuralism.

\section{Superselection and paraparticles}\label{s:para}
In this final Section, our main aim is to complete our revival of Stachel's argument, by showing that IP makes QM an unfixed but permutable theory. Showing this takes some care, for two reasons. First, non-fixity and permutability concern states (viz.~their representing redundantly), while IP concerns states and quantities equally: so there is a gap to be bridged. (In fact, we will note that IP also allows a redundancy in the representation of quantities.)

Second, we will see that there are two very different arguments for QM with IP being an unfixed but permutable theory.  In fact only one of these arguments---the one using paraparticles---serves as a motivation for structuralism.  The other argument, which we consider first, does {\em not} need paraparticles, and corresponds rather to the idea of \emph{superselection}. The first argument is limited to mixed states, which, as we shall see, is why it will not do as a motivation for structuralism.  It may also be overcome, for an advocate of SP can reply to it that her view ``loses nothing'', since every  mixed state of bosons and fermions is perfectly well represented by a density operator obeying SP. Hence our advocacy of the second argument (and our having not hitherto mentioned the first!)---which does use paraparticles, and does apply to pure states.

In Section \ref{subs:IPnotSP}, we will present these two arguments. We adopt a leisurely pace, since the subject can be confusing: not least through too many uses of the word `symmetric' and its cognates! But we will not need details, either technical or conceptual, of superselection or paraparticles.\footnote{For more details, see Messiah \& Greenberg [1964], Hartle \& Taylor [1969], Stolt \& Taylor [1970], French \& Krause ([2006], pp.~131-149), and Caulton \& Butterfield [2010b].} With these arguments, we conclude that QM with IP is an unfixed but permutable theory, but that only the second argument makes QM susceptible---just like GR---to Stachel's argument for structuralism (Section \ref{subs:QGP}). After these generalities, we end with two visualizable examples (Section \ref{subs:examples}). The first, in Section \ref{subs:coins}, illustrates Section \ref{subs:IPnotSP}'s first argument (i.e. superselection between bosonic and fermionic subspaces); and the second, in Section \ref{subs:paras}, illustrates Section \ref{subs:IPnotSP}'s second, successful argument (i.e. paraparticles).

A final preliminary remark. Philosophers of physics tend to ignore paraparticles,\footnote{With the laudable exceptions of French and Huggett.} presumably on the ground that, so far as we know, no fundamental particles are paraparticles. But as mentioned in Section 1, our position---in particular, our revival of Stachel's argument---does not depend on there being such fundamental paraparticles. For our definition of structuralism requires only that there could have been non-individuals, or extrinsic individuating properties.\footnote{Cf.~especially footnote \ref{fn:indessences}.  Besides, although it seems no  fundamental particles
are paraparticles, paraparticle {\em states} are needed in physics, e.g.~in the non-field theory of quarks, although these states only play a role as factors in states that are completely (anti-)symmetric.  Cf.~Caulton \& Butterfield [2010b].}

\subsection{Quantum permutability} \label{subs:IPnotSP}
In this Subsection, we show that QM with IP but not SP is an unfixed but permutable theory. That is: a physical state, understood as a specification of expectation values for all admissible quantities, is represented equally well by each element of an equivalence class of models (i.e.~mathematical states), with the elements related by particle permutations. Since the equivalence class is not a singleton set, these models are under-determined.

As just mentioned, there are two  different arguments for this conclusion. There is no tension between them---both hold good: it is just that giving up SP yields two different sources of under-determination. (But as we admitted, the first may be side-stepped by the advocate of SP, and will not serve as a motivation for structuralism.) \\
\indent (UnderdetMix): The first argument does not require one to consider paraparticles, and concerns only mixed states. It is an under-determination of which density operator represents a mixture of bosonic and fermionic states (or more generally, of symmetry types). It is a typical example of superselection; more specifically, it is the consequence for states of implementing IP by allowing only symmetric quantities. So here we will have superselection of symmetry types:  which will mean---if we set aside paraparticles---that the symmetrised (bosonic) and antisymmetrised (fermionic) subspaces are superselection sectors.  As we shall argue later (Section \ref{subs:QGP}), the fact that this under-determination applies only to mixed states will mean that it will not provide an argument for structuralism.\\
\indent (UnderdetPure): The second argument does involve paraparticles, and amounts to an underdetermination of which ray represents a pure state. That is: for paraparticles, a pure state will be represented not by one ray, but by an equivalence class of them; (in fact, the class of all rays lying in a multi-dimensional irreducible representation of the permutation  group).  To this second argument, the advocate of SP has, we submit, no reply. In particular, no density operator obeying SP can represent such paraparticle states (pure or mixed).

We lay out these arguments in Section \ref{subsubs:qmpermaby}. But both arguments need some preliminaries, first mathematical (Section \ref{subsubs:511}) and then physical (Section \ref{subsubs:512}). These develop a little the earlier Sections' formal material.

\subsubsection{Symmetry types and symmetric operators}\label{subsubs:511}
We begin with the action of the permutation (also known as: symmetric) group $S_n$,  comprising all permutations on $n$ objects, on the $n$-particle assembly's Hilbert space $\mathcal{H} =  \bigotimes_l^n \mathfrak{h}_l$. Recall that $\cal H$ carries a unitary representation of $S_n$: $P: \pi \in S_n \mapsto P(\pi)$. For the case $n = 2$, $S_n$ has just two elements, $S_2 = \{ id, (12) \}$, so that Section \ref{subsubs:SPIP}'s definitions of fully symmetrised, and fully antisymmetrised, vectors simplify: $|\psi \rangle$ is fully symmetrised  iff $P(12) |\psi \rangle = |\psi \rangle$, and $|\psi \rangle$ is fully antisymmetrised  iff $P(12) |\psi \rangle = - |\psi \rangle$. The following facts are readily checked.\\
\indent (i): For the case $n = 2$, these two kinds of vector form orthogonal subspaces, ${\cal S(H)}$ and ${\cal A(H)}$. The projectors onto these subspaces are respectively, $E_{\cal S} := E_{\cal S(H)}:= \frac{1}{2}({\mathbb I} + P(12))$ and $E_{\cal A} := E_{\cal A(H)} := \frac{1}{2}({\mathbb I} - P(12))$.\\
\indent (ii): Again for the case $n = 2$, $|\psi \rangle$ is fully symmetrised  iff its components in any product basis are symmetric, $c_{ij} = c_{ji}$; and similarly, $|\psi \rangle$ is fully antisymmetrised  iff we have $c_{ij} = - c_{ji}$. Then with $d =$ dim($\mathfrak{h}$), we infer that dim(${\cal S(H)}$) = the number of independent components of a symmetric $d \times d$ matrix = $\frac{1}{2}d(d+1)$, and dim(${\cal A(H)}$) = the number of independent components of an antisymmetric $d \times d$ matrix = $\frac{1}{2}d(d-1)$. But dim($\cal H$) = $d^2$. So for $n=2$, $\cal H$ has no further subspaces orthogonal to both ${\cal S(H)}$ and ${\cal A(H)}$. That is: ${\cal H} = {\cal S(H)} \oplus {\cal A(H)}$. In other words, $E_{\cal S} + E_{\cal A} = {\mathbb I}.$\\
\indent (iii): For $n \geq 3$, we similarly define symmetric and antisymmetric subspaces ${\cal S(H)}$ and ${\cal A(H)}$; which are orthogonal. The projectors onto these subspaces are, respectively, $E_{\cal S} := E_{\cal S(H)} := \frac{1}{n!}\sum_{\pi} \; P(\pi)$ and $E_{\cal A} := E_{\cal A(H)} := \frac{1}{n!}\sum_{\pi} \; (-)^{\pi} P(\pi)$; (where $(-)^{\pi}$ is the parity of $\pi$). But for $n \geqslant 3$, $\cal H$ does have further subspaces orthogonal to both ${\cal S(H)}$ and ${\cal A(H)}$. That is, $E_{\cal S} + E_{\cal A} < {\mathbb I}.$ In fact, ${\cal S(H)} \oplus {\cal A(H)}$ is the set of vectors invariant under even permutations; and the orthocomplement $({\cal S(H)} \oplus {\cal A(H)})^{\perp}$, comprising vectors not invariant under even permutations, breaks down into various summands exhibiting various behaviours under $S_n$. {\em These} are the various paraparticle sectors.

These points can be put in the language of group representations. The representation $P: \pi \in S_n \mapsto P(\pi)$ decomposes into sub-representations. On ${\cal S(H)}$, the  representation is utterly faithless: the restriction of any $P(\pi)$ to ${\cal S(H)}$ is just the identity ${\mathbb I}_{{\cal S(H)}}$.\footnote{Incidentally, and in reference to foonote \ref{fn:Huggett}, this is why all quantities, when restricted to $\mathcal{S(H)}$ (or $\mathcal{A(H)}$), are symmetric.} So any orthobasis of ${\cal S(H)}$ decomposes the representation as a sum of one-dimensional irreducible representations: each on the ray of the corresponding  basis element, and each utterly trivial: $P(\pi) = {\mathbb I}: c \in {\mathbb C} \mapsto c \in {\mathbb C}$.  Similarly, on ${\cal A(H)}$: each $P(\pi)$ acts either like the identity or like its negative, according as $\pi$ is even or  odd. Any orthobasis of ${\cal A(H)}$ decomposes the representation as a sum of one-dimensional irreducibles, each ``quasi-trivial'': $P(\pi)  \mapsto (-)^{\pi}{\mathbb I}_{\mathbb C}$. But note that for $n \geq 3$, there is also the orthocomplement $({\cal S(H)} \oplus {\cal A(H)})^{\perp}$: whose various summands are various multi-dimensional irreducible representations of $S_n$.

Now we turn to the action of $S_n$ on operators; and thereby to the representation of states and quantities---and to SP and IP. There is a natural action of $S_n$ on the set of linear operators, $A: {\cal H} \rightarrow {\cal H}$, defined by: for each $\pi \in S_n$, $A \mapsto P(\pi)A P(\pi)^{\dagger}$. Here, we  write $A$ for a general linear operator, so as to cover both a quantity (self-adjoint operator) $Q$ and a density operator $W$. Indeed, this is a representation of $S_n$ on the vector space End($\cal H$) of the operators $A$. Recall that in Section \ref{subsubs:SPIP}, we defined an operator, in particular a quantity $Q$, to be symmetric iff it commutes with all the $P(\pi)$, i.e. iff for all $\pi$, $Q = P(\pi)Q P(\pi)^{\dagger}$, i.e. iff $Q$ is fixed by this action.\footnote{This gives a partial concordance of jargon, despite `symmetric' being a dangerously over-used word in this area! Namely: this usual definition of a symmetric operator corresponds to our (more general) use of `fully symmetric' to mean `fixed by the relevant action of $S_n$'; e.g. in Section \ref{subsubs:SPIP}'s definition of `fully symmetric vector/ray'. Cf. (2) of footnote \ref{jargonremarks} and footnote \ref{jargonremarks2}.} The following facts are readily checked.\\
\indent (i): Any eigenspace of a symmetric operator is invariant under $S_n$ (i.e. under $S_n$'s action on vectors). \\
\indent (ii): For a projector $E$ on $\cal H$: the range ran($E$) $\equiv E({\cal H})$ is invariant under the action of $S_n$ iff $E$ is symmetric. (The leftward implication is trivial; the rightward implication uses the fact that the representation $\pi \mapsto P(\pi)$ is unitary, so that ran($E$) being invariant implies that ran($E$)$^{\perp}$ is also.) So in particular, the projectors onto the symmetric and antisymmetric subspaces, $E_{\cal S}$ and $E_{\cal A}$, are symmetric. More generally, the projector  onto any sub-representation---e.g. a single fully (anti)symmetrised ray in the range of $E_{\cal S}$ or of $E_{\cal A}$, or a multi-dimensional irreducible representation, or a sum of such irreducible representations---is symmetric. \\
\indent (iii):  Taken together, (i) and (ii) imply: the projector onto an eigenspace of any symmetric operator is itself symmetric. In particular, the spectral projectors of any self-adjoint symmetric operator are symmetric. (This also follows from noting that any function of a self-adjoint symmetric operator is symmetric.)\\
\indent (iv): A symmetric operator is block-diagonalized by the complete orthogonal family of projectors  onto the various symmetry sectors. We need not consider the structure of the paraparticle sectors. We can just write the projector onto their sum as $E_{\cal P}$ ($\cal P$ for `paraparticle'). So we have: $E_{\cal P} = {\mathbb I}_{\cal H} - E_{\cal S} - E_{\cal A}$. Now, if $A$ is a symmetric operator, then $[A, E_{\cal S}] = 0 = [A, E_{\cal A}]$. So in the expansion,
\begin{equation}
A = {\mathbb I}_{\cal H} A {\mathbb I}_{\cal H} = (E_{\cal S} + E_{\cal A} + E_{\cal P})A(E_{\cal S} + E_{\cal A} + E_{\cal P}) = E_{\cal S}AE_{\cal S} + E_{\cal A}AE_{\cal A} + E_{\cal P}AE_{\cal P} \;\;
+ \; \mbox{ cross-terms}
\end{equation}
the cross-terms vanish. For each includes either $E_{\cal A}$ or $E_{\cal S}$ (or both) which pass through $A$ to meet $E_{\cal S}$ or $E_{\cal A}$ or $E_{\cal P}$; and $E_{\cal S} \perp E_{\cal A}$, $E_{\cal S} \perp E_{\cal P}$, $E_{\cal A} \perp E_{\cal P}$.

We turn to the analogue, for $S_n$'s action on the space End($\cal H$) of operators, of the projector $E_{\cal S(H)}$ on $\cal H$. Thus we define the {\em symmetriser} $\bf \Sigma$ that sends $A: {\cal H} \rightarrow {\cal H}$ to a symmetric ``cousin'' of $A$:
\begin{equation}
{\bf \Sigma}: \;  \mbox{End}({\cal H}) \ni A  \;\; \mapsto \;\; {\bf \Sigma}(A) \equiv A_{\textrm{sym}}
\;\; := \; \; \frac{1}{n!} \; \sum_{\pi} \; P(\pi)AP(\pi)^{\dagger}.
\label{defSigma}
\end{equation}
The following facts are readily checked.\\
\indent (i): If $A$ is symmetric, $A = {\bf \Sigma}(A)$. Also, symmetrised operators (i.e. operators of the form in eq.~(\ref{defSigma})) are symmetric; so $A$ is symmetric iff $A = {\bf \Sigma}(A)$.\\
\indent (ii):  $\bf \Sigma$ is a projector on End($\cal H$): not just in the sense that it fixes each symmetric $A \in \mbox{End}({\cal H})$, and so is idempotent ${\bf \Sigma}^2 = {\bf \Sigma}$; but also it is self-adjoint once we make End($\cal H$) a complex Hilbert space by defining the sesquilinear form (called `Hilbert-Schmidt') $\langle A_1, A_2 \rangle : = Tr(A_1^{\dagger}A_2)$. (This form is non-degenerate, since $A^{\dagger}A$ is self-adjoint and positive.) That is: one checks that $\langle A_1, {\bf \Sigma}(A_2) \rangle = \langle {\bf \Sigma}(A_1), A_2 \rangle$. To sum up: End($\cal H$)'s subspace of symmetric operators is ran($\bf \Sigma$).\\
\indent (iii): Combining (i) with the block-diagonalization in (iv) above, we deduce: if $A$ is symmetric, then
\begin{equation}
{\bf \Sigma}(A) = A = E_{\cal S}AE_{\cal S} + E_{\cal A}AE_{\cal A} + E_{\cal P}AE_{\cal P}.
\label{SigmaAblock}
\end{equation}
\indent (iv): Symmetrization, i.e. the action of ${\bf \Sigma}$, preserves the trace, being self-adjoint, and being a density operator. But it does not preserve idempotence, or being a  projector. For let $F$ be the projector onto the ray spanned by $|\psi_s\rangle + |\psi_a\rangle$, a linear combination of $|\psi_s\rangle \in {\cal {S(H)}}$ and $|\psi_a\rangle \in {\cal {A(H)}}$. Then ${\bf \Sigma}(F) = E_{\cal S}FE_{\cal S} + E_{\cal A}FE_{\cal A}$ represents a mixture of a bosonic and a fermionic state; (cf. (i) at the start of Section \ref{subsubs:SPIP}).\\
\indent (v): According to Section \ref{subsubs:SPIP}, a bosonic density operator, i.e. a $W$ such that $W = P(\pi)W$ for all $P(\pi)$, is symmetric; and so is a fermionic density operator $W$, i.e. a $W$ such that $W = (-)^{\pi}P(\pi)W$. But the converse fails. That is: a symmetric operator $W$ can have a non-zero component on the paraparticle sector, i.e. $E_{\cal P}WE_{\cal P} \neq 0$.

\subsubsection{Statistics and symmetrization; superselection}\label{subsubs:512}
We can now prove two easy but important results about the action of $\bf \Sigma$; with very similar (one might say: dual) statements and proofs---and a close relation to superselection.\\
\indent  (a): The first concerns the Born-rule statistics, for any quantity, that are prescribed by a symmetrised, i.e. symmetric, density operator. This will show that when working with such density operators (in particular, under SP), there is a redundancy in the representation of physical quantities.\\
\indent  (b): The second concerns the Born-rule statistics, prescribed by any density operator, for a symmetric quantity, i.e. one that obeys IP. This will show that with symmetric quantities, there is a redundancy in the representation of physical states.

Before we state these results, two warnings about the (a)-(b) contrast being different from two other contrasts. First: the (a)-(b) contrast is not just a reflection of the SP-IP contrast. For recall that a density operator's obeying SP is sufficient but not necessary for its being symmetric, i.e. for the assumption of (a). This paper also favours (b) in so far as we emphasize throughout redundancy in the representation of states, not of quantities---as provided by (a); recall from the end of Section \ref{subsubs:perms}, that we take quantities as a `fixed background'.\footnote{But agreed, one could explore this paper's topics, assuming throughout symmetric operators and with quantities redundantly represented. We believe that our appeal to paraparticles to revive Stachel's argument would go through unaffected. But we will not pursue this: we have work enough for one paper!} Second: the (a)-(b) contrast  does {\em not} correspond to our two promised arguments, (UnderdetMix) and (UnderdetPure), for under-determination, which we announced at the start of Section \ref{subs:IPnotSP}. Both (a) and (b) apply equally to (UnderdetMix) and to (UnderdetPure). But as just discussed, both arguments
will stress the under-determination of states, not of quantities.

(a): For any density operator $W$ and any self-adjoint operator $Q$,
\begin{equation}
Tr\left({\bf \Sigma}(W)Q\right) =  Tr\left({\bf \Sigma}(W){\bf \Sigma}(Q)\right) \; \; ; \; \; \mbox{i.e.} \;\;
Tr\left(W_{\textrm{sym}}Q\right) =  Tr\left(W_{\textrm{sym}}Q_{\textrm{sym}}\right)
\label{SigmaOnW}
\end{equation}
In words, using the Born rule: for any (even un-symmetric) quantity $Q$, a symmetric state $W = W_{\textrm{sym}}$ encodes only the statistics of the symmetric ``cousin'' of $Q$. (Agreed, the trace formula encodes directly only $Q$'s expectation value: not the probabilities  for its various eigenvalues. But this statement holds good, i.e. this other information is encoded: consider the spectral projectors of $Q$.)

The proof is simple. Using the linearity and cyclicity of trace, and $W_{\textrm{sym}}$ commuting with all $P(\pi)$, we deduce that the right hand side of eq.~(\ref{SigmaOnW}) is
\begin{eqnarray}
Tr\left(W_{\textrm{sym}}\left(\frac{1}{n!} \; \sum_{\pi} \; P(\pi)QP(\pi)^{\dagger}\right)\right) & = &
\frac{1}{n!} \; \sum_{\pi} \; Tr\left(W_{\textrm{sym}}P(\pi)QP(\pi)^{\dagger}\right) \nonumber \\
&=& \frac{1}{n!} \; \sum_{\pi} \; Tr\left(P(\pi)^{\dagger}W_{\textrm{sym}}P(\pi)Q\right) \nonumber \\
& = & \frac{1}{n!} \; \sum_{\pi} \; Tr\left(W_{\textrm{sym}}Q\right) = \frac{n!}{n!} \; Tr\left(W_{\textrm{sym}}Q\right) \nonumber \\
&\equiv& Tr\left(W_{\textrm{sym}}Q\right).
\label{proof(a)}
\end{eqnarray}
Note that the proof does not depend on $W$'s being a density operator, or $Q$'s being self-adjoint. So mathematically, eq.~(\ref{SigmaOnW}) holds for any linear  operators $W, Q$; only our physical interpretation of it so depends.

(b): For any density operator $W$ and any self-adjoint operator $Q$,
\begin{equation}
Tr\left(W{\bf \Sigma}(Q)\right) =  Tr\left({\bf \Sigma}(W){\bf \Sigma}(Q)\right) \; \; ; \; \; \mbox{i.e.} \;\;
Tr\left(WQ_{\textrm{sym}}\right) =  Tr\left(W_{\textrm{sym}}Q_{\textrm{sym}}\right)
\label{SigmaOnQ}
\end{equation}
In words, using the Born rule: for any (even non-symmetric) density operator $W$, a symmetric quantity  $Q = Q_{\textrm{sym}}$ is given the same statistics as are given by $W_{\textrm{sym}}$, the symmetric ``cousin'' of $W$. (As in (a), one takes spectral projectors so as to infer the statement about all of $Q$'s statistics.)

The proof is parallel to that for (a). The only differences are that we expand $W_{\textrm{sym}}$, not $Q_{\textrm{sym}}$; and we use the fact that $Q_{\textrm{sym}}$, rather than $W_{\textrm{sym}}$, commutes with all $P(\pi)$.

Results (a) and (b) are readily expressed in terms of equivalence relations on quantities and states, defined by $Q_1 \sim Q_2$ iff ${\bf \Sigma}(Q_1) = {\bf \Sigma}(Q_2)$ and $W_1 \sim W_2$ iff ${\bf \Sigma}(W_1) = {\bf \Sigma}(W_2)$. Using eq.~(\ref{SigmaAblock}), we can define the relation $\sim$ on all linear operators $A$, in words: $A_1 \sim A_2$ iff $A_1$ and $A_2$ have the same block-truncation by symmetry sectors. One readily checks that $\sim$ is closed under convex combination, i.e. if $A_1 \sim A_2$ then for $\lambda, \mu \in [0,1], \lambda + \mu = 1$, we have $\lambda A_1 + \mu A_2 \sim A_1$. But it is not closed under superposition of (the vectors that generate) rays. For let $A_1, A_2$ project onto the rays, skew to ${\cal S(H)}$ and ${\cal A(H)}$, generated by $|\psi_s \rangle \pm | \psi_a \rangle$, so that $A_1 \sim A_2$; these vectors can be superposed to give e.g.~$|\psi_s \rangle$, which lies completely within $\mathcal{S(H)}$.

Thus our equations \ref{SigmaOnW} and \ref{SigmaOnQ} are, respectively:
\begin{equation}
Tr({\bf \Sigma}(W)Q_1) = Tr({\bf \Sigma}(W)Q_2)
 \; \; ; \; \; \mbox{and} \;\;
Tr(W_1{\bf \Sigma}(Q)) =  Tr(W_2{\bf \Sigma}(Q)) .
\label{SigmaOnWandQequivce}
\end{equation}
Clearly, these equivalence classes of quantities and states mean there is redundancy of representation. To sum up: (a) if we assume states are symmetric (as in SP), a quantity is represented redundantly by all (self-adjoint) operators with the same image under ${\bf \Sigma}$; and (b) if we assume quantities are symmetric (as in IP), a state is represented redundantly by all (density) operators with the same image under ${\bf \Sigma}$.

{\em Digression: superselection} \quad For completeness, we relate our discussion to {\em superselection}. This is almost always defined as a restriction on the set of quantities, with an ensuing redundancy of the representation of states: so that our (b) is an example. And usually, the restriction on quantities is defined by the existence of a set of operators (called `superselection operators') that commute with all quantities; so that again, taking the permutation operators, our (b) is an example.
 It is also often assumed, if only for simplicity, that the   superselection operators commute and are simultaneously diagonalizable. Of course, this is {\em not} obeyed by the permutation operators: they do not commute! But one can instead take the superselection operators to be the mutually orthogonal projectors  onto the different symmetry sectors: $E_{{\cal S}}$, $E_{{\cal A}}$ and the projectors onto the various paraparticle sectors. In any case, let us now assume that the   superselection operators commute and are simultaneously diagonalizable---just to make the connection between our discussion and elementary superselection as simple as possible. We will also assume that the simultaneous diagonalization of the superselection operators gives a discrete, not continuous, block-decomposition of the state-space.

So we assume that $\cal H$ is a direct sum of subspaces ${\cal K}_{\alpha}$, ${\cal H} = \oplus_{\alpha} \; {\cal K}_{\alpha}$, where the projectors $E_{\alpha}$ onto the ${\cal K}_{\alpha}$ are superselection operators, i.e. commute with all quantities $Q$. Then any $Q$ has zero matrix-elements between different ${\cal K}_{\alpha}$, i.e. is correspondingly block diagonalized. (So for our purposes, the projectors $E_{\alpha}$ are $E_{{\cal S}}$, $E_{{\cal A}}$ and the projectors onto the various paraparticle sectors.) So the analogue of the results, eq.~(\ref{SigmaOnW}) and (\ref{SigmaOnQ}), in (a) and (b) above is that for any density operator $W$ and any quantity $Q$, i.e. self-adjoint operator commuting with all the  $E_{\alpha}$,
\begin{equation}
Tr(WQ) =  Tr(\Sigma_{\alpha} (E_{\alpha} W E_{\alpha})Q) \; .
\label{supereselecOnW}
\end{equation}
The proof is like that of (a) or (b). One applies linearity and cyclicity of trace; and then uses that fact that $[Q, E_{\alpha}] = 0$, so that $\Sigma_{\alpha} (E_{\alpha} Q E_{\alpha}) = Q$.

Finally, it is worth noting as a corollary, the quantum no-signalling theorem for a non-selective L\"uders-rule (i.e. projection postulate) measurement. For in terms of density operators, such a measurement involves a state-transition $W \mapsto \Sigma_{\alpha} (E_{\alpha} W E_{\alpha})$, where the $E_{\alpha}$ are the spectral projectors for the quantity measured on, say, the left-component of a two-component system. Then one expects a quantity $Q$ pertaining to the right-component to commute with all the $E_{\alpha}$---yielding eq.~(\ref{supereselecOnW}): which, thus interpreted, says that the expectation value of any quantity on the right-component is unaffected by a non-selective measurement on the left-component. {\em End of digression}

\subsubsection{The two arguments for quantum permutability}\label{subsubs:qmpermaby}
With these results, especially eq.~(\ref{SigmaOnQ}), we can now easily give our two promised arguments (UnderdetMix) and (UnderdetPure), that quantum theory with IP is an unfixed but permutable theory. For the essential points are already established, as follows.

\indent Non-fixity means that (mathematical, rather than physical) states are not fixed by permutations. Indeed, for the first argument (UnderdetMix): even without paraparticles, vectors, and so rays, that are skew to both ${\cal {S(H)}}$ and ${\cal {A(H)}}$ are not fixed by odd permutations; and so also for (un-symmetrised) density operators that decompose into (projectors onto) such rays. And for (UnderdetPure): with paraparticles, ${\cal {P(H)}}$ is a sum of multi-dimensional irreducible representations of $S_n$; so vectors, and so rays, that are in ${\cal {P(H)}}$ are in general not fixed by a permutation; and so also for density operators that decompose into (projectors onto) such rays.

On the other hand, permutability means that (mathematical) states that are permutes of each other are selected invariantly; i.e.~in quantum theory, they ascribe the same statistics to all quantities. (This follows from  eq.~(\ref{SigmaOnQ}).) Thus, for (UnderdetMix): even without paraparticles, if $W_2 = P(\pi) W_1 P(\pi)^{\dagger}$, then ${\bf \Sigma}(W_2) = {\bf \Sigma}(W_1)$, so that by eq.~(\ref{SigmaOnQ}), $W_1$ and $W_2$ prescribe the same statistics for all symmetric quantities $Q$. Finally, for (UnderdetPure): with paraparticles, exactly the same result applies. To sum up: with IP, ${\bf \Sigma}(W_1) = {\bf \Sigma}(P(\pi) W_1 P(\pi)^{\dagger})$ is the quantum expression of permutability.

We should emphasize that the linear structure of quantum theory implies that for both these under-determinations, of mixed states and pure states, each equivalence class of states is larger than we would classically expect for a permutable theory. Classically, each class consists only of the permutes of any given element, i.e.~state or model. (This applies equally to the space of pure states, and the space containing also mixed states---recall (from just before eq.~(\ref{SigmaOnWandQequivce})) that the $\sim$-classes are closed under convex combination.) But quantum theory's linear structure makes the equivalence classes larger.

To see this for the first argument (UnderdetMix), we need only note that (rays generated by) vectors $| \psi_s \rangle + e^{i \theta} | \psi_a \rangle$, $| \psi_s \rangle \in {\cal S(H)}$ and $| \psi_a \rangle \in {\cal A(H)}$, with various different relative phases $\theta$ yield the same statistics for all symmetric quantities. So the rays are $\sim$-equivalent, i.e.~are in the same $\sim$-class.  But for the second argument (UnderdetPure), the point is more striking: the equivalence class of a pure paraparticle state---that is, of a ray in one of the multi-dimensional irreducible representations---is the \emph{entire} representation, not merely the set of its permutes. That is: the given ray's $\sim$-class contains all the rays in the representation, not just its permutes. This result depends on a simple but crucial theorem we have not yet invoked, viz.~(one form of) Schur's Lemma.  We can state it as follows. Suppose we are given a representation of a group $G$ on a complex vector space, and a linear operator $Q$ that commutes with all representatives of $G$. Then: (i) $Q$ is block-diagonalized by the various inequivalent sub-representations of $G$; and (ii) in the subspace spanned by a set of equivalent irreducible representations, $Q$ is block-decomposed into multiples of the identity, i.e. if $V_i, V_j$ are two equivalent irreducible representations of dimension $m$, then the $(i,j)$-block of $Q$ is $c_{ij}{\mathbb I}_m$, with $c_{ij} \in {\mathbb C}$; (here $V_i$ might be $V_j$, so that also along the block-diagonal, $Q$ is a multiple of the identity). Now we take $G$ as $S_n$, represented on $\cal H$, and $Q$ as any symmetric quantity, and $V_1$, say, as the irreducible representation containing our given pure paraparticle state (indeed, since it is irreducible: generated by the state). Since $Q$ is block-decomposed into multiples of the identity, with its $(1,1)$-block being $c_{11}{\mathbb I}_m$, every ray in $V_1$ gives $c_{11}$ as the expectation value to $Q$. In other words: the initial ray's $\sim$-class contains the whole of $V_1$.

\subsection{Quantum general permutability}\label{subs:QGP}
Quantum mechanics with IP is therefore an unfixed, but permutable theory; accompanied (as such theories are) by the familiar flagrant, but mercifully unphysical, under-determination.  So the analogy between quantum mechanics and general relativity, which Stachel originally claimed, appears to be restored!

But, as mentioned in the preamble to Section \ref{s:para}, there is a problem with our first argument (UnderdetMix).  Recall our modified version of Stachel's argument:

\begin{enumerate}
\item [$1'$.] Structuralism about {unfixed} $T$ is defined by holding general permutability for $T$.
\item [$2'$.] General permutability for {unfixed} $T$ explains/makes palatable the permutability of $T$.
\item [3.] Therefore (by inference to the best explanation): the permutability of $T$ supports structuralism about $T$.
\end{enumerate}

We submit that premise $2'$ cannot be denied: that is, we believe that, when faced with the underdeterminaton of density operators in (UnderdetMix), the best strategy is to take a whole $\sim$-equivalence classes of density operators to represent a single physical state.  This may seem to conflict with the pragmatic approach of simply choosing an arbitrary, or the most convenient, density operator from a $\sim$-class to represent the physical state.  But this approach must admit that other density operators could, in principle, equally well do the job.  (Which density operator is the most convenient is anyway not a fixed matter, though the preferred candidates are usually rays or symmetric density operators.) 

Thus the problem for (UnderdetMix) being a motivation for structuralism lies in premise $1'$.  We claim that premise $1'$ fails in a way that could not be anticipated in a classical context, since the problem relates to the difference between pure and mixed states.

For general permutability to express a commitment to structuralism, we must take it that the relevant feature invariant over a given $\sim$-class is \emph{structure}.  In a classical context, it is enough for this that the equivalence class is generated by permutations of the relevant objects (in this case: the particles; in the case of GR: the spacetime points).  In (UnderdetMix), the $\sim$-classes are certainly generated by permutations of the particles (though, as just explained in the Section \ref{subsubs:qmpermaby}, each class ends up with far more than a single state's permutes).  But the non-singleton $\sim$-classes are populated by entirely mixed states: and we should require non-singleton $\sim$-classes of \emph{pure} states, in order to claim that the permutations are really swapping objects ``under" some invariant structure.  

The point here is not just the general one that a mixed state might be proper, i.e.~ignorance-interpretable---and we should no doubt be wary of any argument seeking to derive ontological consequences from considerations about such states, which after all include an expression of our epistemic situation.  There is also a specific point, which is well illustrated by the quantum coins of Section \ref{subsubs:badarg}.

For these coins, consider the $\sim$-class containing the states $|HT\rangle\langle HT|$ and $|TH\rangle\langle TH|$.  Written out in full, with particle labels included, these two states are $|H\rangle\langle H|_1 \otimes |T\rangle\langle T|_2$ and $|T\rangle\langle T|_1 \otimes |H\rangle\langle H|_2$. It is \emph{very} tempting to think that what differs between these two states is simply which particle is which, and therefore that what is invariant between the two states is the \emph{structure} of the qualitative properties and relations enjoyed by the two particles.   This temptation goes along with another temptation, to conceive of both states as ``genuine" product states.  On this conception, we can interpret e.g.~the state $|H\rangle\langle H|_1 \otimes |T\rangle\langle T|_2$ as attributing the single-particle state $|H\rangle\langle H|$ to particle 1 and $|T\rangle\langle T|$ to particle 2.  But such an interpretation would be available \emph{only if the state were pure}.  The interpretation is not available if the state is mixed, since the class of admissible quantities simply does not support it.  IP demands that all admissible quantities be \emph{symmetric}, so no quantity exists which could support the interpretation that particle 1 was in some definite state (\emph{H}) and particle 2 was in another (\emph{T}).  Therefore, despite the way they are written, in fact the two states do not differ about which particle is which.  The illusion that they differ is brought about by allowing ourselves to write down states in a way (viz.~as rays) that suggests information that goes beyond what is in fact possessed by the physical state.  This extra pseudo-information comprises, of course, those off-diagonal elements of the density operator that connect two superselection sectors, in this case the symmetry sectors.\footnote{In this respect, the mixed states are somewhat analogous to antisymmetric vectors.  These too are not invariant under all permutations (they change sign under odd permutations), but one should not consider the equivalence classes $\{|\psi\rangle, -|\psi\rangle\}$ as representing invariant \emph{structure}.  Rather, the sign of the vector is pseudo-information not relevant to the physical state; it is for this reason that (symmetric and antisymmetric) pure states are better represented by \emph{rays} rather than vectors.}

The case of (UnderdetPure) is much happier. There permutability applies to the {pure}, \emph{and therefore also} to the mixed, states.  We can therefore legitimately take the $\sim$-classes, generated by the particle permutations, to represent invariant \emph{structure}: premise $1'$ in this context is true.  We hold to premise $2'$, as always: the way to restore complete determination of states by the expectation values of physical quantities, as in the theory of distinguishable particles, is to adopt an appropriate many-to-one representation relation between mathematical representations and physical reality, i.e.~between rays and quantum states.  The appropriate strategy involves representing states by irreducible invariant subspaces of $\mathcal{H}$ under the action of $S_n$, which are in general \emph{multi}-dimensional.  This proposal was made by Messiah \& Greenberg [1964], who suggested calling these subspaces \emph{generalised rays}.  Therefore, by inference to the best explanation, we obtain our structuralist conclusion.

\subsection{Examples of quantum permutability}\label{subs:examples}
We end with a brief look at with two easily visualized examples of quantum permutability. The first, in Section \ref{subs:coins}, illustrates Section \ref{subs:IPnotSP}'s first argument, (UnderdetMix), i.e. superselection between bosonic and fermionic subspaces. The second, in Section \ref{subs:paras}, illustrates Section \ref{subs:IPnotSP}'s second argument, (UnderdetPure), i.e. paraparticles.

\subsubsection{Two quantum coins, again}\label{subs:coins}
We return to the quantum coins of Section \ref{subsubs:badarg}.  Without SP, the full four-dimensional Hilbert space has state-vectors of the general form
$$
|\psi\rangle = \tau|HH\rangle + \xi|HT\rangle + \eta|TH\rangle + \zeta|TT\rangle.
$$
However, for the sake of simplicity (and visualizability!)~in this example, let us concentrate on a restricted part of the full Hilbert space.  We will set aside the ``double-result" vectors $|HH\rangle$ and $|TT\rangle$, so we set $\tau = \zeta = 0$.  Our restricted state-vectors now take the general form
$$
|\psi\rangle =  \xi|HT\rangle + \eta|TH\rangle .
$$
With IP in mind, we express $|\psi\rangle$ in a physically more salient basis:
\begin{eqnarray*}
|\psi\rangle &= & \xi\frac{1}{\sqrt{2}}\left[\frac{1}{\sqrt{2}}\left(|HT\rangle + |TH\rangle\right) +\frac{1}{\sqrt{2}}\left(|HT\rangle - |TH\rangle\right)  \right]\\
&& \qquad + \
 \eta\frac{1}{\sqrt{2}}\left[\frac{1}{\sqrt{2}}\left(|HT\rangle + |TH\rangle\right) - \frac{1}{\sqrt{2}}\left(|HT\rangle - |TH\rangle\right)  \right] \\
 &=:&\frac{1}{\sqrt{2}}(\xi + \eta)|\psi_s\rangle \ + \ \frac{1}{\sqrt{2}}(\xi - \eta)|\psi_a\rangle\ .
 \end{eqnarray*}
 where $|\psi_s\rangle$ and $|\psi_a\rangle$ are the fully symmetrised (bosonic) and fully antisymmetrised (fermionic) unit vectors, respectively.

\begin{figure}[h]
   \centering
   \includegraphics[width=\textwidth]{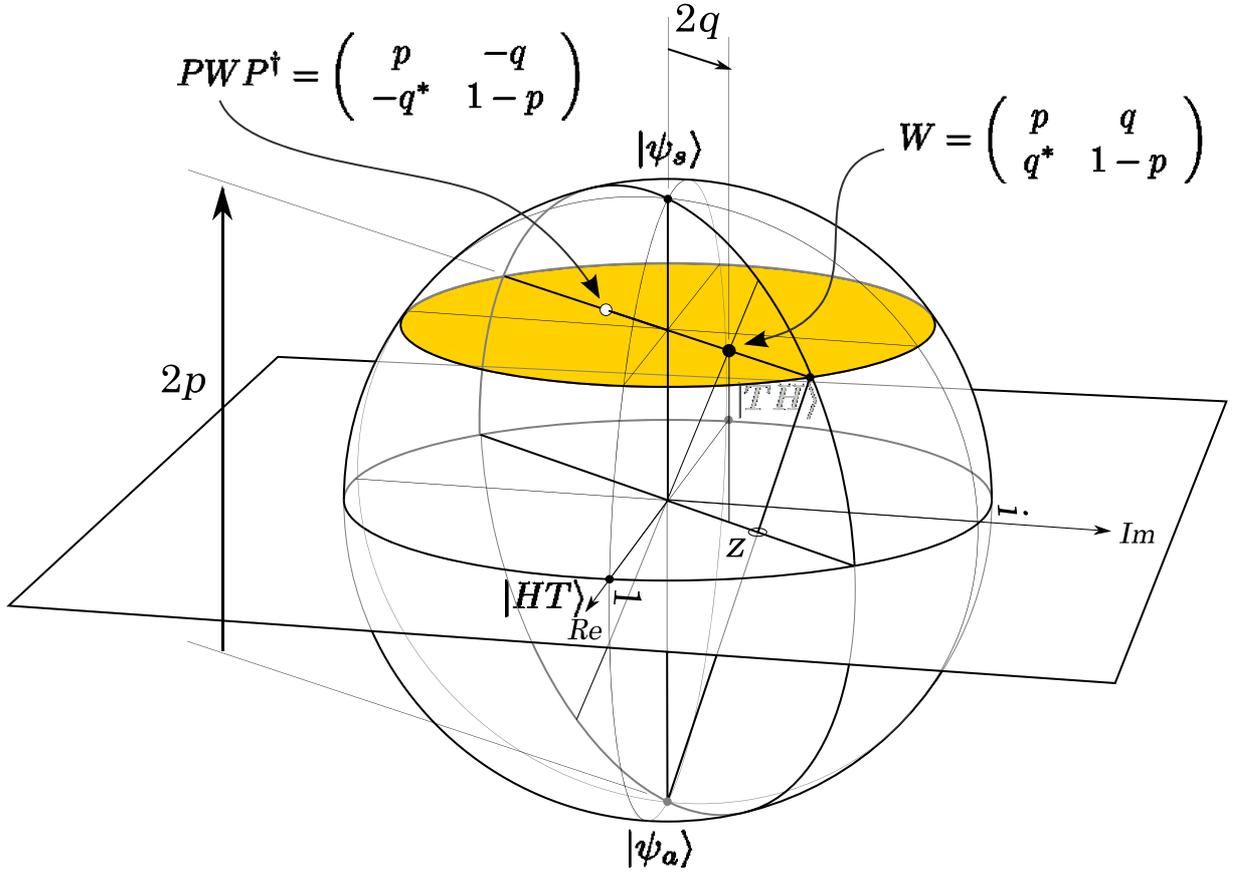}
   \caption{An (interesting) part of the state-space for two ``bosonic coins". \label{fig:Bloch}}
\end{figure}

Since the normalisation of vectors is only a convention, all that really matters for the representation of states is the relative amplitude between $|\psi_s\rangle$ and $|\psi_a\rangle$. Consequently, the state represented by $\psi$ is completely specified by a single complex number, $z: =\frac{\xi - \eta}{\xi+\eta}$.  We can plot the various values of $z$ on the surface of a sphere (the Bloch or Riemann sphere) by stereographic projection;  cf.~Figure \ref{fig:Bloch}.  We set the equator in the complex plane and project from the South pole. So $|\psi_s\rangle$, corresponding to $z = 0$, is plotted onto the North pole; and $|\psi_a\rangle$, corresponding to the circle at $z = \infty$, is plotted onto the South pole.  Thus unit-vectors (equivalently, rays) are plotted one-to-one on the surface of this sphere.  Besides, density operators that are not projectors can also be added to the representation: since they are convex combinations of projectors, we can plot them as points \emph{within} the sphere. 

In the `symmetry basis' $\{|\psi_s\rangle,|\psi_a\rangle\}$, a generic density operator 
has the form $W = 
p|\psi_s\rangle\langle\psi_s| +
(1-p)|\psi_a\rangle\langle\psi_a| +
q|\psi_a\rangle\langle\psi_s| +
q^*|\psi_s\rangle\langle\psi_a|$, where $p \in \mathbb{R}$ and $q \in \mathbb{C}$.  $p$ is the probability of the state being bosonic, and it may readily be checked that 
$
p  = \frac{1}{1+|z|^2}
$.  By similar triangles, it may also be checked (cf.~Figure \ref{fig:Bloch}) that the height, measured from the South pole, of the point representing $W$ is $2p$.  The off-diagonal element $q$ indicates the ``degree of mixedness" of the state $W$: for rays, i.e.~$W =
\frac{1}{1+|z|^2}\left(|\psi_s\rangle +
z|\psi_a\rangle\right)
\left(\langle\psi_s| +
z^*\langle\psi_a|\right)
$, $q$ clearly takes the maximum value $\frac{z}{1 + |z|^2}$, so that $|q|^2 = p(1-p)$; and $W$ is symmetric---that is, a mixture of the bosonic and fermionic states---iff $q=0$.  It may also be checked that the point representing $W$ is vertically projected onto $2q$ in the complex plane.  Thus $p$ and $q$ offer a geometrically natural specification of $W$.

We may also represent $W$ as a $2 \times 2$ matrix, as follows:
$$
W = \left(
\begin{array}{cc}
p & q^*\\ q & 1-p
\end{array}
\right),
\qquad
\mbox{where}
\qquad
p = \frac{1}{1+|z|^2}
\qquad
\mbox{and}
\quad
\begin{array}{c}
\\ |q| \leqslant \sqrt{p(1-p)} \\
\mbox{\small{(with equality for rays)}}
\end{array} .
$$

Quantities are also represented by $2\times 2$ matrices.  In the symmetry basis, IP demands that off-diagonal entries of the symmetric quantities are zero; cf.~eq.~(\ref{SigmaAblock}) in Section \ref{subsubs:512}.  Writing the representation of the flip between the two coins as `$P$', and a generic symmetric quantity as `$Q$',
$$
\langle \psi_a|Q|\psi_s\rangle = -\langle P\psi_a|Q|P\psi_s\rangle
= -\langle \psi_a|P^\dag QP|\psi_s\rangle
= -\langle \psi_a|Q|\psi_s\rangle
$$
(where we have used IP in the last equality); so IP requires that $\langle \psi_a|Q|\psi_s\rangle = 0$.

This means that we obtain {identical} specifications for the expectation value $\langle Q\rangle := Tr(QW)$ with density matrices $W$ which have the same diagonal entries (i.e.~the same value for $p$), no matter what their off-diagonal entries (i.e.~no matter what value for $q$); cf.~eq.~(\ref{SigmaOnQ}) in Section \ref{subsubs:512}.  This is an instance of permutability, specifically (UnderdetMix), discussed in Section \ref{subsubs:qmpermaby}.  General permutability now demands that each {physical} state is represented, not by a \emph{point} in the sphere, but by an entire horizontal \emph{slice} through it, all points of which share the same value for $p$.

\subsubsection{A generalised ray} \label{subs:paras}
Let us be as elementary as possible: we consider three particles, each of which is associated with a two-dimensional Hilbert space; and for the purposes of visualisation, we will ignore the difference (the factor of two!)~between the number of complex and real dimensions.  This gives us eight dimensions for the assembly Hilbert space (which we represent below as real), but we need only focus on three of these.  We choose the orthonormal basis $\{|\alpha\rangle, |\beta\rangle\}$ in the single-particle Hilbert space, and focus on the subspace in the assembly's Hilbert space spanned by the product-state vectors $|\alpha\alpha\beta\rangle, |\alpha\beta\alpha\rangle$ and $|\beta\alpha\alpha\rangle$. Note that this subspace is the span of the $\sim$-equivalence class (cf.~eq.~(\ref{SigmaOnQ}) and ensuing discussion) of each product-state.

\begin{figure}[h]
   \centering
   \includegraphics[width=\textwidth]{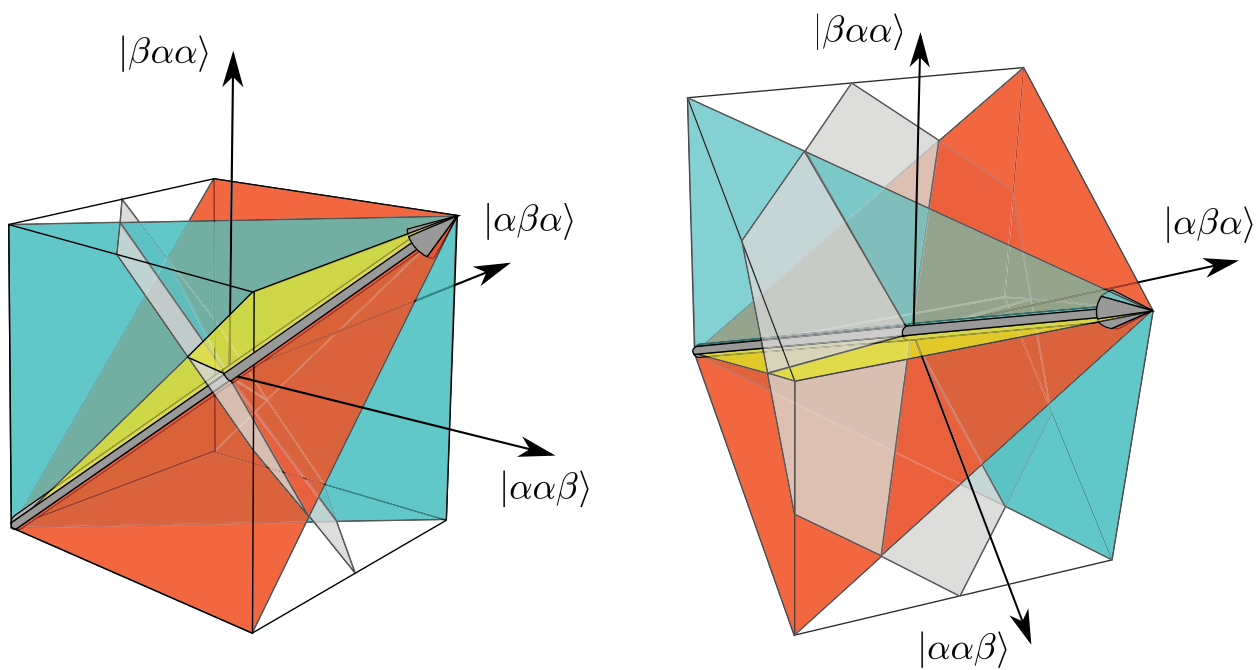}
   \caption{A three-dimensional subspace of a three-particle Hilbert space. There are two irreducible invariant subspaces: the ray (grey arrow), representing a bosonic state; and the plane (light grey hexagon), representing a paraparticle state.\label{fig:paras}}
\end{figure}

This three-dimensional subspace offers a reducible representation of the $3! = 6$ particle permutations. The representation may be displayed rather intuitively in this space, cf.~Figure \ref{fig:paras}.  The effect of each particle \emph{swap} ((12), (23), (13)) is to \emph{reflect} the vectors of the subspace through a plane.  In the figure, the operator which reflects through the $(|\alpha\alpha\beta\rangle, |\alpha\beta\alpha\rangle + \beta\alpha\alpha\rangle)$-plane represents the swap between particles 1 and 2, the $(|\beta\alpha\alpha\rangle, |\alpha\alpha\beta\rangle + \alpha\beta\alpha\rangle)$-plane swaps particles 2 and 3, and the $(|\alpha\beta\alpha\rangle, |\alpha\alpha\beta\rangle + \beta\alpha\alpha\rangle)$-plane swaps particles 1 and 3; the remaining two non-trivial permutations are then composed out of the swaps.

From Section \ref{subsubs:511}, we know that the irreducible representations will correspond to the states of definite symmetry type.  (And since this subspace contains only states in which the $\alpha$ state occurs twice, we know that there will be no fermionic representation, due to Pauli exclusion.)  In fact there are two irreducible representations: one is obvious, it is the ray along which the three reflecting planes intersect; this is the fully symmetric representation and so represents a  bosonic state.

The second irreducible representation  lies perpendicular to all three reflecting planes: it is the plane perpendicular to the fully symmetric ray.  The reader may wish to satisfy him- or herself that no vector on this plane can leave it under any permutation.  This plane---called by Messiah \& Greenberg ([1964]) a \emph{generalised ray}---is the span of the six permutes of any paraparticle ray chosen in this three-dimensional subspace.  Here we have an instance of permutability, in line with (UnderdetPure) in Section \ref{subsubs:qmpermaby}, since the \emph{entire} plane is an equivalence class of indistinguishable states.  General permutability now demands that we take the whole plane to represent a single paraparticle state.

\ \\
{\Large \bf Acknowledgements}

We thank Erik Curiel, Nick Huggett, Stephen Leuenberger, Oliver Pooley, John Stachel and Paul Teller for very helpful comments on a draft version; and seminar audiences in Cambridge, Paris, Amsterdam and Pittsburgh.

\begin{flushright}
Adam Caulton\\
{\em Faculty of Philosophy\\
University of Cambridge\\
CB3 9DA, UK\\
adam.caulton@gmail.com}

\ \\

Jeremy Butterfield\\
{\em Faculty of Philosophy\\
University of Cambridge\\
CB3 9DA, UK\\
jb56@cam.ac.uk}
\end{flushright}

\ \\
{\Large \bf References}
\ \\

Adams, R.~M. [1979]: `Primitive thisness and primitive identity', \emph{Journal of Philosophy} \textbf{76}, pp. 5-26.

Ainsworth, P.~M. [2009]: `Newman's objection', {\em British Journal for the Philosophy of Science}, {\bf 60}, pp. 135-71.

Armstrong, D.~M. [1989]: \textit{A Combinatorial Theory of Possibility}.  Cambridge: Cambridge University Press.

Belot, G. [2003]: `Notes on symmetries', in Brading and Castellani [2003], pp. 393-412.

Belot, G. and Earman, J. [2001]: `Pre-Socratic Quantum Gravity', in C.~Callender and N.~Huggett (eds.),~{\em Physics Meets Philosophy at the Planck Scale}.  Cambridge: Cambridge University Press, pp.~213-55.

Bishop, R.~L. and Goldberg, S.~I. [1968]: \textit{Tensor Analysis on Manifolds.} New York: Dover.

Brading, K. and Castellani, E. [2006]: `Symmetries and invariances in classical physics', in J. Butterfield  and  J. Earman (eds.), \textit{The Handbook of Philosophy of Physics}, North Holland, pp.~1331-67.

Brading, K. and Castellani, E. (eds.) [2003]: \emph{Symmetries in Physics: Philosophical Reflections}, Cambridge: Cambridge University Press.

Butterfield, J.~N. [1987]: 'Substantivalism and determinism', {\em International Studies in the
Philosophy of Science}, {\bf 2}, pp.~10-32.

Butterfield, J.~N. [1989]: `The Hole Truth', \textit{British Journal for the Philosophy of Science} \textbf{40}, pp.~1-28.

Butterfield, J.~N. [1993]: `Interpretation and identity in quantum theory', \emph{Studies in the History and Philosophy of Science} \textbf{24}, pp. 443-76.

Butterfield, J.~N. [2006]: `On symplectic reduction in classical mechanics', in J. Butterfield  and  J. Earman (eds.), \textit{The Handbook of Philosophy of Physics}, North Holland, pp.~1-131.

Carnap, R. [1928]: {\em Der Logische Aufbau der Welt}, translated by R. George [1967]: {\em The Logical Structure of the World}.  Berkeley: University of California Press.

Carnap, R. [1950]: {\em The Logical Foundations of Probability}. Chicago:  Chicago University Press.

Carnap, R. [1966]: {\em The Philosophical Foundations of Physics}.  New York:  Basic Books.

Castellani, E. (ed) [1998]: \emph{Interpreting Bodies: Classical and Quantum Objects in Modern Physics}.  Princeton University Press.

Caulton \& Butterfield [2010a]: `On identity and indiscernibility in logic and metaphysics', submitted to \emph{The British Journal for the Philosophy of Science}.

Caulton \& Butterfield [2010b]: `Paraparticles for philosophers', in preparation.

Da Costa, N. and Krause, D. [2007]: `Logical and Philosophical Remarks on Quasi-set Theory', {\em Logic Journal of the IGPL}, {\bf 15}, pp.~1-20.

Dalla Chiara, M.~L.,~Guintini, R.~and Krause, D. [1998]: `Quasiset Theories for Microobjects: A Comparison', in Castellani [1998], pp.~142-52.

Dieks, D. [1990]: `Quantum Statistics, Identical Particles and Correlations', {\em Synthese} {\bf 82}, pp.~127-55.

Earman, J.  [2010, preprint]: `Understanding permutation invariance in quantum mechanics'.

Earman, J. and Norton, J. D. [1987]: `What Price Spacetime Substantivalism', \textit{British Journal for the Philosophy of Science}, \textbf{38}, pp.~515-25.

Esfeld, M. and Lam, V. [2008]: `Moderate structural realism about space-time'. \textit{Synthese} \textbf{160}, pp.~27-46.

French, S. [1998]: `On the Withering Away of Physical Objects', in {\em
Interpreting Bodies}, ed. E. Castellani, Princeton: University Press, pp.~93-113.

French, S. and Ladyman, J. [2003]: `Remodelling Structural Realism: Quantum Physics and the Metaphysics of Structure', \emph{Synthese}, \textbf{136} (1), pp.~31-56.

French, S. and Krause, D. [2006]: \emph{Identity in Physics}.  Oxford: Oxford University Press.

French, S. and Redhead, M. [1988]: `Quantum physics and the identity of indiscernibles', \emph{British Journal for the Philosophy of Science} \textbf{39}, pp.~233-46.

Gasiorowicz, S. [1974]: \textit{Quantum Physics}. New York: Wiley.

Greenberg, O. W. [2009]: `Generalizations of quantum statistics', in Greenberger, D., Hentschel K., and Weinert, F. (eds.), \emph{Compendium of Quantum Physics}, Berlin: Springer, pp.~255-58.

Hacking, I. [1977]: `The identity of indiscernibles', \emph{Journal of Philosophy} \textbf{72}, pp.~249-56.

Hartle, J.~B. and Taylor, J.~R. [1969]: `Quantum mechanics of paraparticles', \textit{Physical Review} \textbf{178} (5), pp.~2043-51.

Hawking, S.~W. and Ellis, G.~F.~R. [1973]: \textit{The Large Scale Structure of Space-Time}.  Cambridge: Cambridge University Press.

Hilbert, D. and Bernays, P. [1934]:  \emph{Grundlagen der Mathematik}, vol. 1.  Berlin: Springer.

Hoefer, C. [1996]: `The metaphysics of space-time substantivalism', \emph{Journal of Philosophy} \textbf{93}, pp.~5-27.

Huggett, N. [1999]: `On the significance of permutation symmetry', \emph{British Journal for the Philosophy of Science} \textbf{50}, pp.~325-47.

Huggett, N. [2003]: `Quarticles and the identity of indiscernibles', in Brading and Castellani [2003], pp.~239-249.

Huggett, N. and Imbo, T. [2009]: `Indistinguishability', in Greenberger, D., Hentschel K., and Weinert, F. (eds.), \emph{Compendium of Quantum Physics}, Berlin: Springer, pp.~311-7.

Jantzen, B.~C. [2009, preprint]: `No Two Entities Without Identity'.

Kaplan, D. [1966]: `Transworld Heir Lines', in Loux [1979], pp. 88-109.

Ketland, J. [2004]: `Empirical Adequacy and Ramsification', {\em British Journal for the Philosophy of Science}, {\bf 55}, pp.~287-300.

Krause, D. [1992]: `On a quasi-set theory', \emph{Notre Dame Journal of Formal Logic} \textbf{33} (3), pp.~402-11.

Krause, D. and French, S. [2007]: `Quantum sortal predicates', \emph{Synthese} \textbf{143}, pp.~417-30.

Ladyman, J. [1998]: `What is Structural Realism?', {\em Studies in the History and Philosophy of Science}, {\bf 55}, pp.~287-300.

Ladyman, J. [2007a]: `Structural Realism' entry in \textit{The Stanford Encyclopaedia of Philosophy.}

Ladyman, J. [2007b]: `On the identity and diversity of objects in a structure',  \emph{Aristotelian Society Joint Session Supplementary Volume}, {\bf 81}, pp.~23-43.

Lehmkuhl, D. [2008]: `Is spacetime a gravitational field?', in Dieks, D. (ed.), \emph{The Ontology of Spacetime II}, Amsterdam: Elsevier, pp.~83-110.

Leitgeb, H. and Ladyman, J. [2008]: `Discussion Note: Criteria of identity and structuralist ontology',\emph{Philosophica Mathematica}, {\bf 16}, pp.~388-96.

Lewis, D. [1970]: `On the Definition of Theoretical Terms', \emph{Journal of Philosophy} \textbf{67}, pp.~427-46.

Lewis, D. [1986]: {\em On the Plurality of Worlds}. Oxford: Blackwell.

Loux, M. [1979]: {\em The Possible and the Actual: readings in the metaphysics of modality}, New York: Cornell University Press.

Margenau, H. [1944]: `The exclusion principle and its philosophical importance', \emph{Philosophy of Science} \textbf{11}, pp. 187-208.

Massimi, M. and Redhead, M. [2003]: `Weinberg's proof of the spin-statistics theorem', {\em Studies in the History and Philosophy of Modern Physics}, {\bf 34}, pp.~621-50. 

Maudlin, T. [1990]: `Substances and Space-Time: What Aristotle Would Have Said to Einstein', \textit{Studies in the History and Philosophy of Science}, {\bf 21}, pp.~531-61.

Melia, J. and Saatsi, J. [2006]: `Ramseyfication and Theoretical Content', {\em British Journal for the Philosophy of Science}, {\bf 57}, pp.~561-85.

Messiah, A. [1961]: {\em Quantum Mechanics}, vol.~2. Amsterdam: North Holland.

Messiah, A. and Greenberg, O. W. [1964]: `The symmetrization postulate and its experimental foundation', \textit{Physical Review} \textbf{136} (1B), pp.~248-67.

Misner, C., Thorne, K. and Wheeler, J. [1973]: {\em Gravitation}, San Francisco: Freeman. 

Muller, F.~A. and Saunders, S. [2008]: `Discerning fermions', \emph{British Journal for the Philosophy of Science}, \textbf{59}, pp.~499-548.

Muller, F. A. and Seevinck, M. [2009]: `Discerning elementary particles', \emph{Philosophy of Science}, \textbf{76}, pp.~179-200.

Norton, J. [1984]: `How Einstein found his field equations: 1912-1915', {\em Historical Studies in the Physical Sciences}, {\bf 14}, pp.~253-316.

Ohanian and Ruffini [1994]: \emph{Gravitation and Spacetime} (2nd ed.), New York: W.~W.~Norton \& company.

Pauli, W. [1940]: `The connection between spin and statistics', {\em Physical Review}, {\bf 58}, pp.~716-22.

Pooley, O. [2006]: `Points, particles, and structural realism', in D. Rickles, S. French \& J. Saatsi (eds.), \emph{The Structural Foundations of Quantum Gravity}, Oxford: Oxford University Press, pp.~83-120.

Quine, W. V. O. [1960]:  \emph{Word and Object}.  Cambridge: Harvard University Press.

Quine, W. V. O. [1970]:  \emph{Philosophy of Logic}.  Cambridge: Harvard University Press.

Quine, W. V. O. [1976]: `Grades of discriminability', \emph{Journal of Philosophy} \textbf{73}, pp.~113-6.

Rae, I. [2006]: \textit{Quantum Mechanics, 6th ed.} Institute of Physics: Bristol and Philadelphia.

Redhead, M. and Teller, P. [1992]: `Particle Labels and the Theory of Indistinguishable Particles in Quantum Mechanics', \textit{British Journal for the Philosophy of Science} {\bf 43}, pp.~201-218.

Reichenbach, H. [1956]: `The Genidentity of Quantum Particles', in Castellani [1998], pp.~61-72.

Rynasiewicz [1994]: `The lessons of the hole argument', \emph{The British Journal for the Philosophy of Science}, \textbf{45}, pp.~407-36.

Saunders, S. [2003a]: `Physics and Leibniz's Principles', in Brading and Castellani [2003], pp.~289-307.

Saunders, S. [2003b]: `Indiscernibles, covariance and other symmetries: the case for non-reductive relationism', in A. Ashtkar, D. Howard, J. Renn, S. Sarkar and A. Shimony (eds.), \emph{Revisiting the Foundations of Relativistic Physics: Festschrift in Honour of John Stachel}, Amsterdam: Kluwer, pp.~289-307.

Saunders, S. [2006a]: `Are quantum particles objects?', \emph{Analysis} \textbf{66}, pp.~52-63.

Saunders, S. [2006b]: `On the explanation for quantum statistics', \textit{Studies in the History and Philosophy of Modern Physics}, \textbf{37}, pp.~192-211.

Schiff, L.~I. [1968]: \textit{Quantum Mechanics}, third edition. Tokyo: McGraw-Hill.

Schr\"odinger, E. [1984]: `What Is an Elementary Particle?', in Castellani [1998], pp.~197-210.

Stachel, J. [2002]: ```The Relation between Things" versus ``The Things between Relations": The Deeper Meaning of the Hole Argument', in D.~B.~Malament (ed.),~\emph{Reading Natural Philosophy: Essays in the History and Philosophy of Science and Mathematics}, Illinois: Open Court, pp.~231-66.

Stolt, R.~H. and Taylor, J.~R. [1970]: `Classification of paraparticles,' \emph{Physical Review D} \textbf{1} (8), pp.~2226-8.

Teller, P. [1998]: `Quantum Mechanics and Haecceities', in Castellani [1998], pp.~114-41.

Wald, R. M. [1984]: \textit{General Relativity}. Chicago: University of Chicago Press.

Worrall, J. [1989]: `Structural Realism: The Best of Both Worlds?', {\em Dialectica}, {\bf 43} 1-2, pp.~99-124.

\end{document}